\documentclass[journal,10pt]{IEEEtran}
\usepackage[T1]{fontenc}

\usepackage{amsmath}
\usepackage{amssymb}
\usepackage{amsfonts}
\usepackage{graphicx}
\usepackage{epsfig}
\usepackage{subfigure}
\usepackage{psfrag}
\usepackage{cite}
\usepackage{latexsym}
\usepackage{lettrine}
\usepackage{url}
\usepackage{color}
\usepackage{multirow}

\usepackage{verbatim}

\usepackage{algorithm}
\usepackage{algpseudocode}

\usepackage{bm}
\usepackage{array}
\usepackage{hyperref}					% Reference
\hypersetup{colorlinks,%				% Reference color setup	
	linkcolor=blue,%z
	anchorcolor=blue,
    citecolor=blue}

\newcommand{\tr}{\mathrm{tr}}

\begin{document}
\title{Random ISAC Signals Deserve Dedicated Precoding}
\author{
Shihang Lu,~\IEEEmembership{Graduate Student Member,~IEEE}, Fan Liu,~\IEEEmembership{Senior Member,~IEEE}, Fuwang Dong,~\IEEEmembership{Member,~IEEE},\\ Yifeng Xiong,~\IEEEmembership{Member,~IEEE}, Jie Xu,~\IEEEmembership{Senior Member,~IEEE}, Ya-Feng Liu,~\IEEEmembership{Senior Member,~IEEE}, \\and Shi Jin,~\IEEEmembership{Fellow,~IEEE}

\thanks{Part of this paper has been submitted to IEEE International Conference on Acoustics, Speech and Signal Processing
 (ICASSP) 2024 \cite{lushihang2023sensing}. }
\thanks{(\it Corresponding author: Fan Liu.)}
\thanks{S. Lu, F. Liu, and F. Dong are with the School of System Design and Intelligent Manufacturing (SDIM), Southern University of Science and Technology, Shenzhen 518055, China (e-mail: lush2021@mail.sustech.edu.cn, \{liuf6, dongfw\}@sustech.edu.cn).}
\thanks{Y. Xiong is with the School of Information and Communication Engineering, Beijing University of Posts and Telecommunications, Beijing 100876, China (e-mail: yifengxiong@bupt.edu.cn).}
\thanks{J. Xu is with the School of Science and Engineering (SSE) and the Future Network of Intelligence Institute (FNii), The Chinese University of Hong Kong (Shenzhen), Shenzhen 518172, China (e-mail: xujie@cuhk.edu.cn).}
\thanks{Y.-F. Liu is with the State Key Laboratory of Scientific and Engineering Computing, Institute of Computational Mathematics and Scientific/Engineering Computing, Academy of Mathematics and Systems
Science, Chinese Academy of Sciences, Beijing 100190, China (e-mail:
yafliu@lsec.cc.ac.cn).}
\thanks{S. Jin is with National Mobile Communications Research Laboratory, Southeast University, Nanjing, 210096, China (e-mail: jinshi@seu.edu.cn).}
}

\maketitle
\begin{abstract}
Radar systems typically employ well-designed {\textit{deterministic}} signals for target sensing, while integrated sensing and communications (ISAC) systems have to adopt {\textit{random}} signals to convey useful information. This paper analyzes the sensing and ISAC performance relying on random signaling in a multi-antenna system. Towards this end, we define a new sensing performance metric, namely, ergodic linear minimum mean square error (ELMMSE), which characterizes the estimation error averaged over random ISAC signals. Then, we investigate a data-dependent precoding (DDP) scheme to minimize the ELMMSE in sensing-only scenarios, which attains the optimized performance at the cost of high implementation overhead. To reduce the cost, we present an alternative data-independent precoding (DIP) scheme by stochastic gradient projection (SGP). Moreover, we shed light on the optimal structures of both sensing-only DDP and DIP precoders. As a further step, we extend the proposed DDP and DIP approaches to ISAC scenarios, which are solved via a tailored penalty-based alternating optimization algorithm. Our numerical results demonstrate that the proposed DDP and DIP methods achieve substantial performance gains over conventional ISAC signaling schemes that treat the signal sample covariance matrix as deterministic, which proves that {\textit{random ISAC signals deserve dedicated precoding designs}}.

\end{abstract}


\section{Introduction}
\IEEEPARstart{T}{he} sixth generation (6G) of wireless networks has been envisioned as a key enabler for numerous emerging applications, including autonomous driving, smart manufacturing, digital twin, and low-altitude economy \cite{cui2021integrating, Chafii2023CST}, among which \textit{wireless sensing} is anticipated to play a vital role on top of conventional wireless communication functionalities, giving rise to the \textit{Integrated Sensing and Communications (ISAC)} technology \cite{liuan2022survey,liufan2023seventy}. Indeed, the International Telecommunication Union (ITU) has recently approved the global 6G vision, where ISAC was recognized as one of the six key usage scenarios for 6G \cite{ITU2023}.

The basic rationale of 6G ISAC is to deploy ubiquitous wireless sensing capability over 6G network infrastructures, through the shared use of wireless resources (time, frequency, beam, and power) and the hardware platform \cite{liuxiangTSP2020,liufanTSP2018mimo,chenliTSP2021joint}. Provably, ISAC systems may achieve performance gain over individual sensing and communication (S\&C) systems in terms of the resource efficiency, and may potentially reap extra mutual benefits via sensing-assisted communication and communication-assisted sensing \cite{Nguyen2023TSP,ren2023fundamental, MarwaTWC2023,Huahaocheng2023TWC}. To realize these promises, however, novel ISAC signaling strategies capable of simultaneously accomplishing both tasks are indispensable, which are at the core of the practical implementation of 6G ISAC systems \cite{kobayashiTIT2022,xiong2023fundamental}. In a nutshell, the signal design methodologies for ISAC can be generally categorized as sensing-centric, communication-centric, and joint designs \cite{mishra2019SPM}. While the first two approaches aim to conceive an ISAC waveform on the basis of existing sensing or communication signals, e.g., chirp signals for radar or orthogonal frequency division multiplexing (OFDM) signals for communications \cite{keskin2021TSPlimitedfeedback,chengziyang2021hybrid}, the joint design builds an ISAC signal from the ground-up through precoding techniques, in the hope of attaining a scalable performance tradeoff between S\&C \cite{muxidong2022JSAC,Sankar2023twc,liufan2018TSP,liu2021cramer}. Pioneered by \cite{liufan2018TSP,liu2021cramer}, joint ISAC precoding designs are typically formulated into nonlinear optimization problems, where ISAC precoders are conceived by optimizing a communication/sensing performance metric subject to sensing/communication quality-of-service (QoS) constraints. Typical figures-of-merit include achievable rate, symbol error rate (SER), Cram\'er-Rao bound (CRB), singal-to-interference-plus-noise ratio (SINR), etc \cite{songxianxin2023TSP,mengkt2023TWC,liurang2021JSTSP}.   

Currently, the ISAC design based on 5G New Radio (NR) primarily exploits physical layer reference signals to realize sensing \cite{wei20225g}. In a typical NR frame structure, the reference signals account for only about $10\%$ of the resources in the time-frequency domain \cite{lin20215g}. In view of this, it is necessary to reuse the $90\%$ of time-frequency resources occupied by random communication data signals to realize sensing, thus improving both sensing performance and resource efficiency. Towards that end, ISAC signals must be \textit{random} due to the shared use of data signals for S\&C. More precisely, ISAC signals are supposed to be randomly realized from certain codebooks, such as Gaussian codebook or discrete Quadrature Amplitude Modulation (QAM)/Phase-Shift Keying Modulation (PSK) alphabets \cite{zhangyumeng2023input}. To maximize the achievable communication rate, the information-bearing signals are required to be ``as random as possible'' \cite{xiong2023fundamental,xiong2023generalized}. As an example, Gaussian signals that achieve the capacity of Gaussian channels possess the maximum degree of randomness, in the sense that it has the maximum entropy under an average power constraint \cite{cover1999elements}. In contrast to that, radar sensing systems usually prefer deterministic signals having favorable ambiguity properties, e.g., phase-coded sequences with low range sidelobes \cite{stoica2007probing,bekkerman2006target}, in which case the randomness of the ISAC signal may potentially degrade the sensing performance. This contradictory conflict between S\&C is consistent with both the engineers' experience and S\&C signal processing theory, leading to the fundamental \textit{Deterministic-Random Tradeoff (DRT)} in ISAC systems \cite{xiong2023fundamental,xiong2023generalized}. Recently, the DRT has been theoretically depicted in Gaussian channels \cite{xiong2023fundamental}, by analyzing the Pareto frontier between the sensing CRB and achievable communication rate in a point-to-point ISAC system. More relevant to this work, it has been shown that signal randomness may significantly degrade the favorable ambiguity properties of OFDM ISAC systems \cite{du2023reshaping}.

Although existing schemes are well-designed by sophisticated methods, most of them overlook the random nature of ISAC signals. For instance, many previous studies treat the sample covariance matrix of ISAC signals as deterministic \cite{liu2021cramer,lu2022performance,liu_Yafeng2022joint}, i.e., as equivalent to the statistical covariance matrix, under the assumption that the data frame is sufficiently long. In such a case, one may concentrate solely on designing the ISAC precoder without the need of accounting for the impact brought by the random data. However, such an approximation may not be reliable in practical time-sensitive ISAC scenarios, particularly for ISAC base stations (BSs) equipped with massive antennas. The reason is that the sample covariance matrix converges to its statistical counterpart only when the frame length is much larger than the array size, as will be shown in Sec. \ref{sec2}. This imposes unaffordably huge computational and signal processing overheads for sensing. More severely, unlike conventional radar systems, the state-of-the-art BS may be unable to accumulate massive raw sensory data due to its limited caching capabilities. Consequently, the ISAC BS has to perform sensing by leveraging short-frame observations, where the law of large numbers no longer holds and the data randomness cannot be eliminated. To ensure reliable sensing performance, one has to conceive dedicated precoders to account for the signal randomness, which motivates the study of this paper.

To address the unique challenge arising in ISAC systems due to random signaling, this paper investigates the dedicated precoding designs while considering {\textit {sensing with random signals}} in multi-antenna ISAC systems, where the transmitter aims at estimating the target impulse response (TIR) matrix while communicating with users by emitting information-bearing Gaussian signals. Against this background, we summarize the contributions of our work as follows:
\begin{itemize}
    \item We commence by revisiting the conventional linear minimum mean squared error (LMMSE) estimation through deterministic orthogonal training signals. Then, we define a new metric, namely, ergodic LMMSE (ELMMSE) for sensing with random signals, which is the estimation error averaged over random signal realizations. Moreover, we reveal that the deterministic LMMSE is merely a lower bound of the ELMMSE due to Jensen's inequality, which characterizes the theoretical sensing performance loss due to random signaling. It is worth noting that this conclusion not only applies to scenarios under LMMSE and ELMMSE but generally holds for any convex cost functions for sensing.
    \item We develop novel precoding designs to minimize the ELMMSE for sensing-only scenarios with Gaussian codebooks, namely, data-dependent precoding (DDP) and data-independent precoding (DIP), where the precoder changes adaptively based on the instantaneous realization of Gaussian data, and remains unchanged for all data samples, respectively. We show that the sensing-only DDP problem can be solved in closed form despite its non-convexity and that its DIP counterpart can be solved via the classical stochastic gradient projection (SGP) and momentum-based SGP (MB-SGP) algorithms. To gain more insights into the ELMMSE minimization problem, we unveil the optimal structures of both precoders.
    \item We extend the DDP and DIP strategies to ISAC scenarios by imposing a communication rate constraint, and propose a penalty-based alternating optimization (AO) approach to tackle the non-convex problems. To reduce the complexity of solving DIP problems, we further simplify the stochastic optimization problems by exploiting their high signal-to-ratio (SNR) approximations.
    \item We provide extensive numerical results by comparing with conventional deterministic ISAC precoding designs, which illustrate that both DDP and DIP schemes attain favorable sensing performance under Gaussian codebooks while satisfying the communication requirement. This convincingly suggests that {\it random ISAC signals deserve dedicated precoding designs}, especially when the ISAC BS cannot accumulate sufficient raw sensory data. 
\end{itemize}

The remainder of this paper is organized as follows. Sec. \ref{sec2} introduces the system model of the considered ISAC system and the corresponding performance metrics. Sec. \ref{Sec_III} elaborates on the dedicated DDP and DIP precoding designs for sensing-only scenarios with random signaling. Sec. \ref{Sec_ISAC} generalizes the DDP and DIP framework to ISAC scenarios. Sec. \ref{Sec_V} provides simulation results to validate the performance of the proposed precoders. Finally, Sec. \ref{Sec_VI} concludes the paper.

% {\it Notations}: In this paper, lowercase and uppercase boldface letters refer to vectors and matrices, respectively. $\mathrm{diag}(\bm{A})$ is a diagonal matrix with the entries of $\bm{A}$ on its main diagonal, $[\bm{a}]_k$ means the $k$-th element of $\bm{a}$ and $[a]^{+} = \max(a,0)$. Transpose, conjugate transpose, trace, and Euclidean norm of matrix $\bm{A}$ are denoted by $\bm{A}^{(t)}$, $\bm{A}^{H}$, $\tr(\bm{A})$, and $\|\bm{A}\|$ respectively. $\mathbb{E}_{\bm{A}}(\cdot)$ denotes the expectation regarding a certain random martix $\bm{A}$. $\mathcal{CN}(0,1)$ denotes the circularly symmetric complex normal distribution with zero mean.  

\section{System Model} \label{sec2}

%As illustrated in Fig. \ref{SystemModel}, 
We consider a mono-static multiple-input multiple-output (MIMO) ISAC BS with $N_T$ transmit antennas and $N_R$ receive antennas, which is serving a communication user (CU) with $N_u$ receive antennas while detecting one or multiple targets. Target sensing is implemented over a coherent processing interval consisting of $L$ snapshots. The sensing and communication receive signal models are expressed as  
\begin{subequations}\label{linear_model}
    \begin{align}
        \bm{Y}_c =\bm{H}_c \bm{X}+\bm{Z}_c, \label{linear_comm_model} \\
        \bm{Y}_s =\bm{H}_s \bm{X}+\bm{Z}_s,         \label{linear_sens_model}
    \end{align}
\end{subequations}  
where $\bm{Y}_c \in \mathbb{C}^{N_u \times L }$ denotes the received signal matrix at the CU receiver and $\bm{Y}_s \in \mathbb{C}^{N_R \times L }$ denotes the echoes at the BS sensing receiver, $\bm{H}_c \in \mathbb{C}^{N_u \times N_T}$ represents the point-to-point MIMO channel and $\bm{H}_s \in \mathbb{C}^{N_R \times N_T}$ is the TIR matrix to be estimated\footnote{The TIR matrix may take various forms under different target models. Without loss of generality, this paper will not focus on exploiting the specific structure of $\bm{H}_s$. Readers are referred to \cite{tangbo2019TSP,tangbo2010TSP} for more details.}, $\bm{Z}_c \in \mathbb{C}^{N_u \times L}$ and $\bm{Z}_s \in \mathbb{C}^{N_R \times L}$ denote the additive noise matrix with each entry following $\mathcal{CN}(0,\sigma_c^2)$ and $\mathcal{CN}(0,\sigma_s^2)$, and $\bm{X} \in \mathbb{C}^{N_T \times L }$ denotes the ISAC signal matrix. Moreover, we assume that the TIR matrix follows a wide sense stationary random process with a statistical correlation matrix $\bm{R}_H = \mathbb{E}[\bm{H}_s^H \bm{H}_s]$. Below we commence by revisiting conventional sensing with deterministic signals, followed by sensing and ISAC transmission with random signals.

% \begin{figure}[!t]
% 	\centering
% 	\includegraphics[scale=0.3]{fig/BF_Gaussian.pdf}
% 	\caption{MIMO ISAC system using Gaussian signals.}
%         \label{SystemModel}
% \end{figure} 

\subsection{Sensing with Deterministic Training Signals}\label{II-A}
Let $\bm{S}_D \in \mathbb{C}^{N_T \times L}$ denote a {\it deterministic} training signal satisfying $(1/L)\bm{S}_D\bm{S}_D^H = \bm{I}_{N_T}$. The sensing signal matrix is 
\begin{align}
    \bm{X} = \bm{W} \bm{S}_D,
\end{align}
where $\bm{W}\in \mathbb{C}^{N_T \times N_T}$ represents the precoding matrix to be optimized. Usually, sensing systems aim to minimize detection/estimation errors with respect to some practical constraints (e.g., transmit power), which may be modeled as
\begin{align}\label{General_model_D}
		\min_{\bm{W} \in \mathcal{A}} ~ f(\bm{W} ; \bm{S}_D),
\end{align}
where $\mathcal{A}$ denotes the feasible region of the optimization variable $\bm{W}$. Here, $f(\cdot)$ denotes a (potentially non-convex) loss function of radar sensing\cite{liu2021cramer,liufan2018TSP}, e.g., estimation error. Given a transmit power budget $P$, we may constrain the feasible region as
\begin{align}
    \mathcal{A} \triangleq \left\{\bm{W} \in \mathbb{C}^{N_T \times N_T}  \mid \| \bm{W} \|_F^2 \le P \right\},
\end{align}
in which case problem \eqref{General_model_D} becomes a deterministic optimization problem over a convex set $\mathcal{A}$. 

There are several well-established algorithms for solving problem \eqref{General_model_D}, such as the successive convex approximation (SCA) algorithm and the gradient projection algorithm \cite{boyd2004convex}. In many cases, it may be worth exploring the structure of problem  \eqref{General_model_D}, which would admit a closed-form solution under specific conditions \cite{liu2021cramer,liufan2018TSP}. To facilitate our study, we consider an important application example of \eqref{General_model_D} with a closed-form solution, namely the LMMSE-optimal precoding.

{\it Example (LMMSE-Optimal Precoding):} The celebrated LMMSE estimator of $\bm{H}_s$ is \cite{biguesh2006training}
\begin{align}\label{Estimator_LMMSE}
\hat{\bm{H}}_s=\bm{Y}_s\left(\bm{X}^H \bm{R}_H\bm{X}+\sigma_s^2 N_R \bm{I}_{L}\right)^{-1} \bm{X}^H \bm{R}_H,
\end{align} 
which results in an estimation error of \cite{biguesh2006training}
\begin{align}\label{Cond_MMSE}
{f}(\bm{W} ; \bm{S}_D) &\overset{\quad}{=} \mathrm{tr}\Big[\Big(\bm{R}_H^{-1} + \frac{1}{\sigma_s^2 N_R}\bm{X}\bm{X}^H\Big)^{-1}\Big]\nonumber \\
&\overset{\tiny(a\tiny)}{=} \mathrm{tr}\Big[\Big(\bm{R}_H^{-1} + \frac{L}{\sigma_s^2 N_R} \bm{W}\bm{W}^H\Big)^{-1}\Big],
\end{align}
where $(a)$ holds due to the use of the orthogonal training signal matrix satisfying $\bm{S}_D\bm{S}_D^H = L\bm{I}_{N_T}$\footnote{Note that this assumption is not suitable in the scenarios when $L < N_T$ due to the rank-deficient nature of the signal, e.g., in massive MIMO or accumulation-limited scenarios.}. 
Accordingly, the classical LMMSE-oriented precoding design is to solve the {\it deterministic} optimization problem:
\begin{align}\label{LMMSE_Determinstic}
    \min_{\bm{W} \in \mathcal{A}} ~ {J}_{\mathsf{LMMSE}} \triangleq \mathrm{tr}\Big[\Big(\bm{R}_H^{-1} + \frac{L}{\sigma_s^2 N_R} \bm{W}\bm{W}^H\Big)^{-1}\Big],
\end{align}
which is non-convex despite the feasible region being a convex set. Let $\bm{R}_H = \bm{Q}\bm{\varLambda}\bm{Q}^H$ denote the eigenvalue decomposition (EVD) of $\bm{R}_H$. The optimal precoding matrix is known to be the following {\it water-filling} solution \cite{biguesh2006training}:
\begin{align}\label{LMMSE_Determinstic_Opt_W}
    \bm{W}_{\mathsf{WF}} = \sqrt{\frac{\sigma_s^2{N_R}}{L}} \bm{Q}\left[ \left(\mu_0 \bm{I}_{N_T} - \bm{\varLambda}^{-1}\right)^{+} \right]^{\frac{1}{2}},
\end{align}
where $(x)^{+} = \max (x,0)$ and $\mu_0$ is a constant (commonly known as ``water level'') such that $\|\bm{W}_{\mathsf{WF}}\|_F^2 = P$. 

%\noindent
\textbf{Remark 1.} By emitting the deterministic training signal $\bm{S}_D$, one may always obtain the LMMSE-optimal precoding matrix in \eqref{LMMSE_Determinstic_Opt_W}. However, things become different for ISAC systems, since the transmitted signal should be random to convey useful communication information. In what follows, we elaborate on the general precoding framework while taking the randomness into account.

\begin{figure}[t!]
    \centering
    \includegraphics[width=0.4\textwidth]{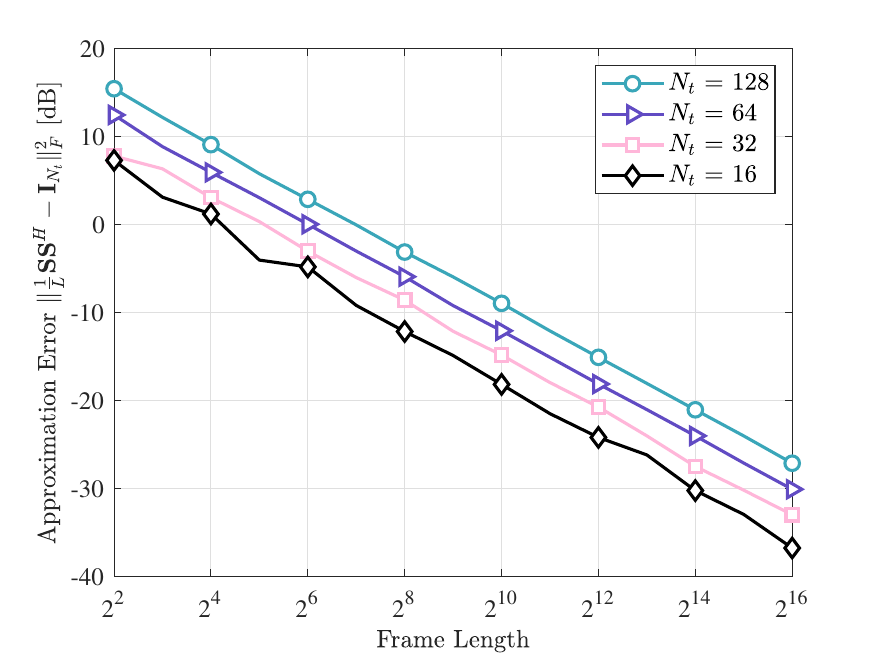}
    \caption{The approximation error of \eqref{SampleMatrix_Appro} versus the frame length.}
    \label{VarMatrix_Appro_Error}
\end{figure}

\subsection{Sensing with Random Signals}

In sharp contrast to radar systems, ISAC systems have to employ random communication signals for sensing. In this paper, we consider Gaussian signaling for ISAC systems, which is known to achieve the capacity of point-to-point Gaussian channels \cite{cover1999elements}. Let $\bm{S} = [\bm{s}(1), \dots, \bm{s}(L)] \in \mathbb{C} ^{N_T \times L}$ denote the transmitted random ISAC signal, where each column is independent and identically distributed (i.i.d.) and follows the circularly symmetric complex Gaussian (CSCG) distribution with zero mean and covariance $\bm{I}_{N_T}$, i.e., $\bm{s}(\ell) \sim \mathcal{CN}(\mathbf{0}, \bm{I}_{N_T})$. It is worth noting that the following assumption is commonly seen in the existing works \cite{liu2021cramer,lu2022performance,liu_Yafeng2022joint},
\begin{align}\label{SampleMatrix_Appro}
    \frac{1}{L} \bm{S}\bm{S}^H \approx \mathbb{E}[\bm{s}(\ell)\bm{s}^H(\ell)] = \bm{I}_{N_T}, \ell = 1,2,\dots,L, 
\end{align} 
where $L \gg N_T$  is assumed such that the approximation holds. 

We show the approximation error of \eqref{SampleMatrix_Appro} by simulation in Fig. \ref{VarMatrix_Appro_Error}. It is observed that to achieve a negligible approximation error, a sufficient accumulation of data samples is necessary. For instance, to keep the approximation error below $-20$ dB for $N_T = 32$, an accumulation of approximately $2^{12} = 4096$ samples is required. This demonstrates that the approximation in \eqref{SampleMatrix_Appro} may not always be accurate in general, especially when the BS is unable to accumulate sufficiently long frames. 

To this end, one needs to take the randomness of the Gaussian signals into account when designing the precoding matrix. More severely, the sensing performance may be randomly varying due to the random signaling. As a consequence, it may not be appropriate to use the conventional estimation metric (e.g., the LMMSE in Sec. \ref{II-A}) relying on the instantaneous realization of signals. To characterize the sensing performance in this practical scenario, we define an {\it ergodic} performance loss\footnote{It is noted that another ergodic sensing performance metric has been well-defined in \cite{xiong2023fundamental}, namely the Miller-Chang type CRB, taking the expectation over all realizations of ISAC signals, which characterizes the estimation error bound under random ISAC signaling.}, as detailed below.

Denote the random ISAC signal matrix by $\bm{S}$ and the loss function of radar sensing by $f(\bm{W}; \bm{S})$. The average sensing performance is defined as $\mathbb{E}_{\bm{S}}[f(\bm{W}; \bm{S})]$, which can be treated as an ergodic sensing metric, namely, a time average over random realizations of $\bm{S}$. In this paper, we consider an ergodic estimation error metric accounting for the signal randomness, namely ELMMSE, expressed as
\begin{align}
    {J}_{\mathsf{ELMMSE}} = \mathbb{E}_{\bm{S}} \left\{\mathrm{tr}\Big[\Big(\bm{R}_H^{-1} +\frac{1}{\sigma_s^2 N_R} \bm{W}\bm{S}\bm{S}^H\bm{W}^H\Big)^{-1}\Big] \right\}.
\end{align}
Accordingly, a general optimization problem for dedicated precoding design towards Gaussian signals is formulated as 
\begin{align}\label{General_Model_Rand}
    \min_{\bm{W} \in \mathcal{A}} ~ \mathbb{E}_{\bm{S}}[f(\bm{W} ; \bm{S})].
\end{align}
Throughout the paper, we focus on the LMMSE loss function by letting $f(\bm{W}; \bm{S}) = \mathrm{tr}[(\bm{R}_H^{-1} +\frac{1}{\sigma_s^2 N_R} \bm{W}\bm{S}\bm{S}^H\bm{W}^H)^{-1}]$. 

\setlength{\parskip}{0.2em}

\textbf{Remark 2.}
Note that problem \eqref{General_Model_Rand} is generally a stochastic optimization problem. Accordingly, the precoding matrix design is non-trivial since the objective function of \eqref{General_Model_Rand} contains expectation, which is distinctly different from problem \eqref{General_model_D}. Below we elaborate on a pair of precoding schemes based on \eqref{General_Model_Rand}, which are with different levels of complexity. %We find that there are mainly two different methods to solve such a stochastic optimization problem with a convex feasible region: one involves obtaining the expectation, e.g., by sampling approximation or asymptotic formulation of the expectation, such that the problem \eqref{General_Model_Rand} can be recast into a deterministic optimization problem; the other revolves the utilization of established stochastic optimization techniques like SGP \cite{li2019convergence}. 

%Therefore, solving problem \eqref{General_Model_Rand} offers a pair of precoding schemes: 

{\textbf{1) DDP Scheme:}} The ELMMSE can be minimized by minimizing the LMMSE for each instance of $\bm{S}$, since the transmitter perfectly knows the data samples due to the mono-static assumption. Therefore, $\bm{W}$ can be designed depending on each realization of $\bm{S}$, leading to a data-dependent precoding scheme where $\bm{W}$ changes adaptively based on the instantaneous value of $\bm{S}$. The DDP scheme may attain the minimum ELMMSE at a high cost of complexity in real-time implementation in practice.

\textbf{2) DIP Scheme:} To reduce the implementation overhead of DDP, an alternative scheme is to find a deterministic precoder $\bm{W}$ that is independent of the signal realization, namely, the data-independent precoder. A data-independent precoder can be optimized offline by generating random training samples based on Gaussian codebooks, which achieves a favorable performance-complexity tradeoff.

Following the philosophy of both DDP and DIP, we will introduce sensing-only precoding schemes for Gaussian signals in Sec. \ref{Sec_III}. 

\subsection{ISAC Precoding with Gaussian Signals}
Recall that $\bm{S}$ is the capacity-achieving Gaussian signal matrix. Therefore, the achievable communication rate (in bps/Hz) of the point-to-point MIMO channel is \footnote{To attain the achievable rate, the communication codeword needs to be sufficiently long and thus spans a large number of coherent processing intervals. In this paper, the instantaneous rate is considered for the communication performance by assuming that the communication channel is flat fading over all intervals, while the ergodic rate is left for our future work. Moreover, the logarithm function is with the base of $2$, denoted by $\log(\cdot)$.}
\begin{align}\label{GaussiaN_Rate}
    R(\bm{W})  = \log \det \left(  \bm{I}_{N_u} + \sigma_c^{-2} \bm{H}_{c} \bm{W} \bm{W}^{H}\bm{H}_{c}^{H} \right).
\end{align}
Without loss of generality, we assume that the ISAC transmitter has the perfect channel state information (CSI) of $\bm{H}_{c}$, and designate the imperfect CSI case as our future work. 

We are now ready to introduce the dedicated precoding optimization problem with random ISAC signals, which is formulated as 
\begin{align}\label{General_Model_ISAC}
		\min_{\bm{W} \in \mathcal{A}} ~  \mathbb{E}_{\bm{S}}[f(\bm{W} ; \bm{S})] ~~
		\mathrm{s.t.} ~ R(\bm{W}) \ge R_0, 
\end{align}
where $R_0$ represents the required communication rate threshold. We will elaborate on dedicated DDP and DIP schemes for ISAC precoding with Gaussian signals in Sec. \ref{Sec_ISAC}. 

\section{Sensing-Only Precoding with Gaussian Signals}\label{Sec_III}
In this section, we investigate the performance boundaries of sensing with random signals by exploring various precoding methods in sensing-only scenarios. This exploration also holds potential advantages for future ISAC deployment, implying that certain sensing tasks could be accomplished by reusing legacy communication signals. Furthermore, the design insights offered by the proposed precoding methods in sensing-only scenarios will be extended to ISAC scenarios, as elaborated in Sec. \ref{Sec_ISAC}. Firstly, we demonstrate the degradation in the sensing performance due to Gaussian signals in comparison to classical deterministic training signals, by Proposition 1 below.

% % \noindent
% % \textbf{\underline{\emph{Proposition}} 1.}
% \textbf{Proposition 1.}
% Gaussian signals lead to ${J}_{\mathsf{ELMMSE}}$ that is no lower than ${J}_{\mathsf{LMMSE}}$ based on the deterministic signals, i.e., the deterministic LMMSE serves as a lower bound of the ELMMSE.

\textbf{Proposition 1.}
Gaussian signals lead to ${J}_{\mathsf{ELMMSE}}$ that is no lower than ${J}_{\mathsf{LMMSE}}$ based on the deterministic signals, i.e., the deterministic LMMSE serves as a lower bound of the ELMMSE.

\textbf{{\emph{Proof}}:}
  Applying Jensen's inequality to ELMMSE immediately yields
    \begin{align}\label{Jensen}
    {J}_{\mathsf{ELMMSE}} & \overset{(a)}{\ge} \mathrm{tr}\Big[\big(\bm{R}_H^{-1} + \frac{1}{\sigma_s^2 N_R} \mathbb{E}_{\bm{S}}\big\{\bm{WS}\bm{S}^H\bm{W}^H\big\}\big)^{-1}\Big]
    \nonumber\\
    &\overset{(b)}{=} \mathrm{tr}\Big[\big(\bm{R}_H^{-1} + \frac{L}{\sigma_s^2 N_R}\bm{W}\bm{W}^H\big)^{-1}\Big]{=} {J}_{\mathsf{LMMSE}},
    \end{align}  
    where $(a)$ holds due to the convexity of $f(\bm{W}; \bm{S})$ with respect to $\bm{WS}\bm{S}^H\bm{W}^H$, and $(b)$ holds, merely by meeting either of the following conditions:

\noindent

\textbf{Condition 1:} If the precoding matrix $\bm{W}$ is independent to the transmit signal matrix $\bm{S}$, the equality $(b)$ holds if $\mathbb{E}_{\bm{S}}[\bm{SS}^H] = L \bm{I}_{N_T}$.

\noindent
\textbf{Condition 2:} If the precoding matrix $\bm{W}$ is dependent on the transmit signal matrix $\bm{S}$, the equality $(b)$ still holds if $\bm{S}$ is a deterministic orthogonal signal such that $\mathbb{E}_{\bm{S}}[\bm{SS}^H] = \bm{S}\bm{S}^H = L\bm{I}_{N_T}$.

To this end, we complete the proof.\hfill $\blacksquare$

% \textbf{{\emph{Proof}}:}
%   Applying Jensen's inequality to ELMMSE immediately yields
%     \begin{align}\label{Jensen}
%     {J}_{\mathsf{ELMMSE}} & \overset{(a)}{\ge} \mathrm{tr}\Big[\big(\bm{R}_H^{-1} + \frac{1}{\sigma_s^2 N_R} \mathbb{E}_{\bm{X}}\big\{\bm{X}\bm{X}^H\big\}\big)^{-1}\Big] \nonumber\\
%     &\overset{(b)}{=} \mathrm{tr}\Big[\big(\bm{R}_H^{-1} + \frac{1}{\sigma_s^2 N_R}\bm{X}\bm{X}^H\big)^{-1}\Big]{=} {J}_{\mathsf{LMMSE}},
%     \end{align}  
%     %     \begin{align}\label{Jensen}
%     % {J}_{\mathsf{ELMMSE}} & \overset{(a)}{\ge} \mathrm{tr}\Big[\Big(\bm{R}_H^{-1} + \frac{1}{\sigma_s^2 N_R} \mathbb{E}_{\bm{S}} \big\{\bm{W}\bm{S}\bm{S}^H\bm{W}^H\big\}\Big)^{-1}\Big]\nonumber \\
%     % &\overset{(b)}{=}\mathrm{tr}\Big[\big(\bm{R}_H^{-1} + \frac{1}{\sigma_s^2 N_R} \mathbb{E}_{\bm{X}}\big\{\bm{X}\bm{X}^H\big\}\big)^{-1}\Big] \nonumber\\
%     % % &\overset{(b)}{=}\mathrm{tr}\Big[\Big(\bm{R}_H^{-1} + \frac{L}{\sigma_s^2 N_R} \bm{W}\bm{W}^H\Big)^{-1}\Big] \nonumber\\
%     % &\overset{(c)}{=} {J}_{\mathsf{LMMSE}},
%     % \end{align}     
%     % where $(a)$ holds due to the convexity of $f(\bm{W} ; \bm{S})$ with respect to $\bm{X}\bm{X}^H$, and $(b)$ holds following the fact that $\mathbb{E}_{\bm{S}}[\bm{S}\bm{S}^H] = L\bm{I}_{N_T}$ \cite{Wishart_Moment}. This completes the proof. \hfill $\blacksquare$
%     where $(a)$ holds due to the convexity of $f(\bm{W} ; \bm{S})$ with respect to $\bm{X}\bm{X}^H$, and $(b)$ holds when $\bm{X}$ is deterministic. This completes the proof. \hfill $\blacksquare$

\begin{figure}[!t]
	\centering
	\includegraphics[width=0.40\textwidth]{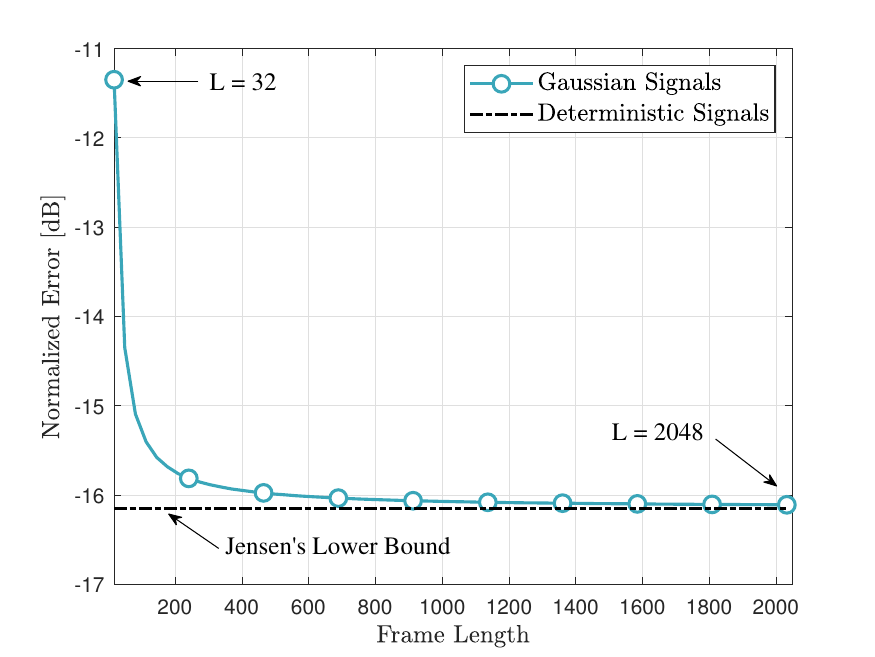}
	\caption{The frame-length-asymptotic performance of Gaussian signals is assessed under the conditions of $N_T = N_R = 32$. The total transmit power is fixed at $16$ dBm, ensuring that the factor of impacting the sensing performance is solely the frame length $L$.}
        \label{IncreasingL}
\end{figure}

%\noindent
\textbf{Remark 3.}
We notice that Jensen's inequality in \eqref{Jensen} tends to be tight with $L$ going to infinity. In light of Proposition 1, we show the asymptotic performance of random signals in Fig. \ref{IncreasingL}, by utilizing water-filling precoding given in \eqref{LMMSE_Determinstic_Opt_W}. It is clearly observed that the assumption in \eqref{SampleMatrix_Appro} is not always reliable especially when the frame length is small. It also justifies the use of ${J}_{\mathsf{ELMMSE}}$ instead of ${J}_{\mathsf{LMMSE}}$ as the performance metric under random signaling, since minimizing ${J}_{\mathsf{LMMSE}}$ only minimizes a lower bound of ${J}_{\mathsf{ELMMSE}}$ instead of the exact estimation error. This observation motivates us to present efficient ELMMSE-oriented precoding designs.

In the subsequent subsections, we first propose a DDP scheme aimed at minimizing ${J}_{\mathsf{ELMMSE}}$, which achieves optimal sensing performance through a closed-form solution but comes at the cost of considerable computational complexity, as the DDP problem has to be solved for every instance of $\bm{S}$. To reduce the complexity, we further present a DIP scheme and propose an SGP algorithm for ELMMSE minimization, which can be trained offline by locally generated signal samples. Finally, we derive the asymptotic formulation of ELMMSE in the high-SNR regime and translate the corresponding ELMMSE-minimization problem into a deterministic optimization problem.

\vspace{-0.5em}
\subsection{DDP: A Closed-Form Solution}
By the law of large numbers, the ELMMSE may be expressed as  
\vspace{-0.5em}
    \begin{align}\label{Sample_Appro_LMMSE}
    \mathbb{E}_{\bm{S}}[f(\bm{W} ; \bm{S})] = \lim_{N\to\infty}\frac{1}{N} \sum_{n=0}^{N-1} {f}(\bm{W};\bm{S}_n),
\end{align}     
where ${f}(\bm{W};\bm{S}_n) = \mathrm{tr}[(\bm{R}_H^{-1} + \frac{1}{\sigma_s^2 N_R} \bm{W}\bm{S}_n\bm{S}_n^H\bm{W}^H)^{-1}]$ and $\bm{S}_n$ denotes the $n$-th Gaussian data realization. Note that in the mono-static sensing setup, each $\bm{S}_n$ is known to both the transmitter and the sensing receiver. Consequently, the DDP precoding matrix $\bm{W}$ can be designed as a function of $\bm{S}$ across all data realization indices $n$, denoted as $\bm{W}(\bm{S}_n)$\footnote{We note that the proposed DDP strategy significantly differs from the classical symbol-level precoding (SLP) exploiting the constructive interference (CI) \cite{liurang2021JSTSP}, in terms of their respective motivations, objective functions, constraints, and design methodologies.}. Therefore, using \eqref{Sample_Appro_LMMSE}, we approximate problem \eqref{General_Model_Rand} as:
\begin{align}\label{LMMSE_Sam}
		\min_{\{ \bm{W}(\bm{S}_n)\in \mathcal{A}\}_{n=1}^{N}} ~ \frac{1}{N} \sum_{n=0}^{N-1} {f}(\bm{W}(\bm{S}_n); \bm{S}_n).
\end{align}    
Hence, the precoding matrix can be successively optimized for each given $\bm{S}_n$ across different data realizations, which decomposes \eqref{LMMSE_Sam} into $N$ parallel deterministic sub-problems:
\begin{align}\label{LMMSE_SubProblem}
		\min_{\bm{W}_n\in \mathcal{A}} ~ f(\bm{W}_n; \bm{S}_n) 
		 , 
\end{align}     
where we define $\bm{W}_n \triangleq \bm{W}(\bm{S}_n)$ for notational simplicity. Notice that problem \eqref{LMMSE_SubProblem} is inherently non-convex. To seek for the optimal solution, we first introduce below a useful lemma on a trace inequality, followed by demonstrating a closed-form solution with Theorem 1.

% \noindent
\textbf{Lemma 1.}
Suppose $\bm{A}$ and $\bm{B}$ are $K \times K $ positive semidefinite Hermitian matrices. Denote their eigen-decompositions by $\bm{A} = \bm{U}_A \bm{\Sigma}_A^{\downarrow} \bm{U}_A^H $ and $\bm{B} = \bm{U}_B \bm{\Sigma}_B^{\downarrow} \bm{U}_B^H $, respectively, where $\bm{\Sigma}_A^{\downarrow} = \mathrm{diag}([\alpha_1,\alpha_2,\dots,\alpha_{K} ])$ and $\bm{\Sigma}_B^{\downarrow} = \mathrm{diag}([\beta_1,\beta_2,\dots,\beta_{K} ])$, such that $\alpha_1 \ge\alpha_2 \ge \dots \ge \alpha_K$ and $\beta_1 \ge \beta_2 \ge\dots \ge \beta_K$. Then the following trace inequality 
\begin{align}
    \mathrm{tr}\big[(\bm{A}+\bm{B})^{-1})\big] &\ge \mathrm{tr}\big[(\bm{\Sigma}_A^{\downarrow}+\bm{P} \bm{\Sigma}_B^{\downarrow} \bm{P})^{-1})\big] \nonumber \\ 
    &= \sum_{i=1}^{K} \frac{1}{\alpha_i + \beta_{K-i+1}}\nonumber
        % \mathrm{tr}\big[(\bm{A}+\bm{B})^{-1})\big] \ge \sum_{i=1}^{K} \frac{1}{\alpha_i + \beta_{K-i+1}}\nonumber
\end{align}
holds, and the equality holds if and only if $\bm{U}_B = \bm{U}_A \bm{P}$, where is a permutation matrix, expressed as
\begin{equation}\label{permutation_mat}
\bm{P}=\left[\begin{array}{cccc}
0 & 0 & \cdots & 1 \\
0 & \cdots & 1 & 0 \\
\vdots & \vdots & \vdots & \vdots \\
1 & 0 & \cdots & 0
\end{array}\right],
\end{equation}
and $\bm{P}$ determines the arrangement of the eigenvalues of $\bm{A}$.

\textbf{{\emph{Proof}}:}
    Please refer to \cite[Appendix A]{tang2011waveform}. \hfill $\blacksquare$

% \noindent
\textbf{Theorem 1.}
Let $\bm{S}_n \triangleq \bm{U}_{\bm{S}_n} \bm{\varSigma}_{\bm{S}_n}  \bm{V}_{\bm{S}_n}^H$ denote the singular value decomposition (SVD) of $\bm{S}_n$. The optimal solution of problem \eqref{LMMSE_SubProblem} is a modified {\it water-filling} solution, expressed as
\begin{align}\label{Opt_DDP}
    \bm{W}_{n}^{\mathsf{opt}} = \bm{Q}\big[ ( \mu_n \bm{\varTheta}_n^{\frac{1}{2}} - \bm{B}_n )^{+} \big]^{\frac{1}{2}} \bm{P}\bm{U}_{\bm{S}_n}^H,~ n = 1, \dots, N,
\end{align}
where $\bm{\varTheta}_n = \frac{1}{\sigma_s^2 N_R} \bm{P}\bm{\varSigma}_{\bm{S}_n}\bm{\varSigma}_{\bm{S}_n}^{T}\bm{P}$, $\bm{B}_n = (\bm{\varLambda}\bm{\varTheta}_n )^{-1}$, $\bm{P}$ is a permutation matrix given by \eqref{permutation_mat}, and $\mu_n$ is a constant selected to satisfy the transmit power constraint $\| \bm{W}_{n}^{\mathsf{opt}} \|_F^2 = P$.

% \textbf{{\emph{Proof}}:} Please refer to Appendix \ref{Theorem_1}.\hfill $\blacksquare$

\textbf{{\emph{Proof}}:}
In light of Lemma 1, it holds immediately that ${f}(\bm{W}_n; \bm{S}_n)$ attains its minimum if and only if 1) $\bm{R}_H^{-1} + (1/\sigma_s^2 N_R)\bm{W}\bm{S}_n\bm{S}_n^H\bm{W}^H$
is diagonal, 2) the eigenvalues of $\bm{R}_H^{-1}$ are sorted in a decreasing order, and 3) the eigenvalues of $(1/\sigma_s^2 N_R) \bm{W}\bm{S}_n\bm{S}_n^H\bm{W}^H$ are sorted in an increasing order. To this end, let $\bm{R}_H^{-1} = \bm{Q}(\bm{\varLambda}^{\uparrow})^{-1}\bm{Q}^H$, where $\bm{\varLambda}^{\uparrow} = \mathrm{diag}([{\lambda}_1,{\lambda}_2,\dots,{\lambda}_{N_T} ]) $ and ${\lambda}_1 \le{\lambda}_2 \le \dots \le {\lambda}_{N_T}$. 
We use $\bm{S}_n \triangleq \bm{U}_{\bm{S}_n} \bm{\varSigma}_{\bm{S}_n}  \bm{V}_{\bm{S}_n}^H$ to denote the SVD of $\bm{S}_n$. By constructing $\bm{W}_n^{\mathrm{opt}} = \bm{Q}\bm{\varLambda}_{\bm{W}_n}\bm{P}\bm{U}_{\bm{S}_n}^H$ where $\bm{\varLambda}_{\bm{W}_n}$ is a diagonal matrix to be optimized. The detailed derivations are analogous to that of \cite[Sec. III-A]{tang2011waveform} and thus omitted here for brevity, thereby completing the proof. \hfill $\blacksquare$

By solving a sequence of parallel sub-problems \eqref{LMMSE_SubProblem} with closed-form solutions, we formulate a set of DDP matrices tailored for each $\bm{S}_n$, expressed by
\begin{align}
    \mathcal{W} = \{ \bm{W}_1^{\mathsf{opt}}, \bm{W}_2^{\mathsf{opt}}, \dots,\bm{W}_N^{\mathsf{opt}}\}. 
\end{align}
Accordingly, the ELMMSE under the DDP scheme may be calculated as
\begin{align}
{J}_{\mathsf{ELMMSE}}^{\mathsf{DDP}}  = \lim_{N\to\infty}\frac{1}{N} \sum_{n=0}^{N-1} f(\bm{W}^{\mathsf{opt}}_n ; \bm{S}_n).
\end{align}    

%\noindent
\textbf{Remark 4.}
The DDP approach attains the minimum ELMMSE of problem \eqref{LMMSE_Sam}. Nevertheless, the design of $\mathcal{W}$ must be conducted sequentially for various realizations of transmitted Gaussian signals, leading to substantially large computational complexity in practical scenarios. This motivates us to design a DIP precoder that remains unchanged for all data realizations, as detailed below.

\subsection{DIP: An SGP-Based Precoding Method}
To conceive a data-independent precoder, we resort to the stochastic gradient descent (SGD) algorithm, which yields a solution with only one or mini-batch samples at each iteration, reducing computational complexity. We first derive the gradient of ${f}(\bm{W};\bm{S})$ at given point $\bm{W}^{\prime}$, expressed as
\begin{align}\label{SGD_Gradient}
\nabla{f}(\bm{W}^{\prime};\bm{S}) = - \frac{1}{\sigma_s^2 N_R} \bm{\varDelta}^{-2}\bm{W}^{\prime} \bm{S}\bm{S}^H, 
\end{align}     
where $\bm{\varDelta} = \bm{R}_H^{-1} + (1/ \sigma_s^2 N_R) \bm{W}^{\prime}\bm{S}\bm{S}^H(\bm{W}^{\prime})^H$. Given the $r$-th iteration point $\bm{W}^{(r)}$, the precoding matrix $\bm{W}$ is updated by
\begin{align}\label{BF_SGDupdate}
 \bm{W}^{(r+1)} \leftarrow \bm{W}^{(r)} - \eta^{(r)} \hat{\nabla} {f}(\bm{W}^{(r)}),
\end{align}     
where $\eta^{(r)}$ denotes the step size (also termed as ``learning rate'') and $\hat{\nabla} {f}(\bm{W}^{(r)})$ is the gradient based on the locally generated mini-batch of Gaussian samples $\mathcal{D}^{(r)}$, given by
\begin{align}\label{min_batch gradient}
\hat{\nabla} {f}(\bm{W}^{(r)}) = \frac{1}{|\mathcal{D}^{(r)}|} \sum_{\bm{S} \in \mathcal{D}^{(r)}} {\nabla} {f}(\bm{W}^{(r)}; \bm{S}).
\end{align}     

Generally, as long as the iterations remain stable, employing a large constant step size $\eta^{(r)}$ facilitates rapid convergence, but only to a large neighborhood of the optimal solution. To increase the accuracy, it becomes necessary to decrease the step size $\eta^{(r)}$. Indeed, the convergence of the SGD algorithm critically hinges on the positive step size $\eta^{(r)}$. For non-convex problems, the most popular step size policy in practice is the decaying step size \cite{liuan2019stochastic}, where the sufficient conditions of convergence guarantee are that $\{\eta^{(r)}\}_{r=1}^{\infty}$ is a deterministic sequence of non-negative numbers satisfying $\sum_{r=1}^{\infty}\eta^{(r)} = \infty~,\sum_{r=1}^{\infty}(\eta^{(r)})^2 < \infty$ \cite{li2019convergence}. In this paper, we opt for $\eta^{(r)} = a/(b+r)$, where $a$ and $b$ are constants chosen to facilitate the convergence. Additionally, a larger $|\mathcal{D}^{(r)}|$ may potentially improve the performance at the cost of computing a greater number of local gradients. 

\begin{algorithm}[!t]
    \caption{SGP Algorithm for Solving \eqref{General_Model_Rand}.}
    \label{PSGD_Alg}
    \begin{algorithmic}[1]
    \Require
    $\bm{R}_H, P, N, \sigma^{2}_{s} , N_T,N_R,r_{\mathrm{max}},\epsilon$.
    \Ensure
    $\bm{W}_{\mathsf{SGP}}$.
    \State Initialize $r = 1$ and $\bm{W}^{(1)}$.
    \State Initialize the Gaussian signal set $\mathcal{S} = \{\bm{S}_1,\bm{S}_2, \ldots , \bm{S}_N\}$.
    \Repeat
    \State Generate $\mathcal{D}^{(r)}$ and calculate $\hat{\nabla}{f}(\bm{W}^{(r)},\bm{S}^{(r)})$.
    \State Update $\bm{W}^{(r+1)} \leftarrow  \mathsf{Proj}_{\mathcal{A}}\big(\bm{W}^{(r)} - \eta^{(r)} \hat{\nabla}{f}(\bm{W}^{(r)})\big) $.
    % \State Take projection $\bm{W}^{(r+1)} \gets \mathsf{Proj}_{\mathcal{A}}\big(\bm{W}^{(r+1)}\big)$.
    \State Update $r=r+1$.
    \Until the increase of the objective value is below $\epsilon$ or $r=r_{\mathrm{max}}$.
    \end{algorithmic}
\end{algorithm}

To maintain compliance with the transmit power budget, we propose to utilize the SGP algorithm, which projects the solution $\bm{W}^{(r+1)}$ onto the feasible region $\mathcal{A}$ after completing a SGD iteration. The projection operator is 
\begin{align}\label{Proj}
\mathsf{Proj}_{\mathcal{A}}( \bm{W})=\left\{\begin{array}{l}
            \bm{W},~ \mathrm{if}~\bm{W} \in \mathcal{A}, \\
            \bm{W}\sqrt{\frac{P}{\|\bm{W}\|_F^2}},~ \mathrm{otherwise}. 
        \end{array}\right.
\end{align}     
Building upon the iteration outlined in \eqref{BF_SGDupdate} and the projection step detailed in \eqref{Proj}, we are now ready to introduce the proposed SGP approach for solving problem \eqref{General_Model_Rand} in Algorithm \ref{PSGD_Alg}. Let $\bm{W}_{\mathsf{SGP}}$ denote the output of Algorithm \ref{PSGD_Alg}. The corresponding ELMMSE under the DIP scheme is
\begin{align}
    {J}_{\mathsf{ELMMSE}}^{\mathsf{DIP}}  = \lim_{N\to\infty}\frac{1}{N} \sum_{n=0}^{N-1} {f}(\bm{W}_{\mathsf{SGP}}; \bm{S}_n).
\end{align}

Moreover, it is known that the popular adaptive moment estimation (Adam) method may help accelerate the convergence of non-convex optimization problems \cite{kingma2014adam,li2019convergence,book2021kkt}. Specifically, Adam estimates the first moment (i.e., the mean $\hat{\nabla}{f}(\bm{W}^{(r)})$) and the second moment (i.e., the uncentered variance $\hat{\nabla}{f}(\bm{W}^{(r)}) \odot \hat{\nabla}{f}(\bm{W}^{(r)})$ with ``$\odot$'' being element-wise multiplication operator) of the gradients by computing the moving averages of the gradients and their squared counterparts, respectively. Then, it utilizes these estimates as momentum to tune the learning rates for each parameter, enabling adaptive adjustments in the parameter space based on the historical performance of gradients and thus accelerating convergence speed \cite{book2021kkt}. However, the classical Adam algorithm mentioned above is only applicable to real-valued optimization problems, since the uncentered variance $\hat{\nabla}{f}(\bm{W}^{(r)}) \odot \hat{\nabla}{f}(\bm{W}^{(r)})$ in complex-variable contexts is not equal to its counterpart with real variables. Therefore, the Adam cannot be harnessed directly into complex-variable situations in our considered ELMMSE minimization problems.

To tackle the above struggling issues, we have established a novel momentum-based SGP (MB-SGP) algorithm, which is a general extension concerning complex-variable optimization problems based on the popular Adam method that works mostly for real-variable optimization problems \cite{kingma2014adam}. Motivated by \cite{book2021kkt}, we propose to keep the momentum terms while slightly modifying the Adam for complex gradients. The key idea here is to leverage $\|\hat{\nabla}{f}(\bm{W}^{(r)})\|_F^2$ to update the second moment of the gradients, which can be treated as the summed uncentered variance of all elements of the complex gradients. To this end, the proposed MB-SGP iteration format reads
\begin{align}\label{MB_SGP_UpdateW}
     \bm{W}^{(r+1)} \leftarrow \mathsf{Proj}_{\mathcal{A}} \left(\bm{W}^{(r)}-\frac{\eta^{(r)}}{\sqrt{\widehat{{v}}^{(r+1)}}+\varepsilon_0} \widehat{\bm{M}}^{(r+1)}\right),
\end{align}
where $\varepsilon_0$ denotes a small value to avoid infinity and
\begin{align}
    \widehat{\bm{M}}^{(r+1)} &= \frac{1}{1-\beta_1^{r}}{\bm{M}}^{(r+1)},~\widehat{{v}}^{(r+1)} = \frac{1}{1-\beta_2^{r}}{{v}}^{(r+1)},  
    \nonumber\\ 
    {\bm{M}}^{(r+1)} &= \beta_1{\bm{M}}^{(r)} + (1-\beta_1) \hat{\nabla}{f}(\bm{W}^{(r)}), \nonumber\\
    {{v}}^{(r+1)} &= \beta_2{{v}}^{(r)} + (1-\beta_2) \|\hat{\nabla}f(\bm{W}^{(r)})\|_F^2. 
    \nonumber
\end{align}
Here, $\widehat{\bm{m}}^{(r+1)}$ and $\widehat{\bm{v}}^{(r+1)}$ represent the ``normalized'' first- and second-moment gradient information by hyperparameters $\beta_1$ and $\beta_2$, respectively. For clarity, we summarize the MB-SGP algorithm in Algorithm \ref{MB_SGD_Alg}. We will illustrate the superiority of the proposed MB-SGP over the SGP algorithm in Sec. \ref{Sec_V}.

\begin{algorithm}[!t]
    \caption{MB-SGP Algorithm for Solving \eqref{General_Model_Rand}.}
    \label{MB_SGD_Alg}
    \begin{algorithmic}[1]
    \Require
    $\bm{R}_H, P, N, \sigma^{2}_{s} , N_T,N_R,r_{\mathrm{max}},\epsilon, \beta_1, \beta_2,\varepsilon_0$.
    \Ensure
    $\bm{W}_{\mathsf{MB-SGP}}$.
    \State Initialize $r = 1$, $\bm{M}^{(1)} = \bm{0}$, ${v}^{(1)} = 0$, and $\bm{W}^{(1)}$.
    \State Initialize the Gaussian signal set $\mathcal{S} = \{\bm{S}_1,\bm{S}_2, \ldots , \bm{S}_N\}$.
    \Repeat
    \State Generate $\mathcal{D}^{(r)}$ and calculate $\hat{\nabla}{f}(\bm{W}^{(r)},\bm{S}^{(r)})$.
    \State Update ${\bm{M}}^{(r+1)} = \beta_1{\bm{M}}^{(r)} + (1-\beta_1) \hat{\nabla}{f}(\bm{W}^{(r)})$.
    \State Update ${{v}}^{(r+1)} = \beta_2{{v}}^{(r)} + (1-\beta_2) \|\hat{\nabla}f(\bm{W}^{(r)})\|_F^2$. 
    \State Update $\widehat{\bm{M}}^{(r+1)} =1/(1-\beta_1^{r}){\bm{M}}^{(r+1)}$
    \State Update $\widehat{{v}}^{(r+1)} = 1/(1-\beta_2^{r}){{v}}^{(r+1)}$.
    \State Update $\bm{W}^{(r+1)}$ by \eqref{MB_SGP_UpdateW}.
    \State Update $r=r+1$.
    \Until the increase of the objective value is below $\epsilon$ or $r=r_{\mathrm{max}}$.
    \end{algorithmic}
\end{algorithm}

We now analyze the complexity of Algorithm \ref{PSGD_Alg} and Algorithm \ref{MB_SGD_Alg}. Let $\mathcal{I}_1$ denote the iteration number to reach convergence. At $r$-th iteration, the SGP algorithm needs to compute the stochastic gradient of $\bm{W} \in \mathbb{C}^{N_T \times N_T}$ with a given batch size $|\mathcal{D}^{(r)}| = D$ by \eqref{SGD_Gradient} and \eqref{min_batch gradient}. The highest cost of the gradient calculation is the inverse operation with a complexity of $\mathcal{O}(N_T^3)$ in \eqref{SGD_Gradient}, and it results in the complexity at the order $\mathcal{O}(D N_T^3)$ in \eqref{min_batch gradient}. Moreover, there is a projection operation with a complexity of $N_t^2$ by \eqref{Proj}. Thus, it requires a complexity of $\mathcal{O}\big(\mathcal{I}_1(D N_T^3+N_T^2)\big)$ in Algorithm \ref{PSGD_Alg}. As for the MB-SGP algorithm, it requires an additional step to update the second-moment information of the gradients with a complexity of $\mathcal{O}(N_T^2)$, resulting in total complexity of $\mathcal{O}\big(\mathcal{I}_1(DN_T^3+2N_T^2)\big)$.

\subsection{DIP in the High-SNR Regime}

To gain more insights into the structure of $\bm{W}_{\mathsf{SGP}}$, we derive an asymptotic formulation of the objective function $\mathbb{E}_{\bm{S}}[f(\bm{W} ; \bm{S})]$ in the high-SNR regime, which is detailed by Lemma 2 below.

\textbf{Lemma 2.}
Let $L>N_T$. In the high-SNR regime, namely when $P/(\sigma_s^2 N_R)$ is sufficiently large, a valid asymptotic formulation of $\mathbb{E}_{\bm{S}}[f(\bm{W} ; \bm{S})]$ is expressed by
\begin{multline}\label{ELMMSE_HSNR_Appro}
\mathbb{E}_{\bm{S}}[f(\bm{W} ; \bm{S})] \approx  \tilde{f}(\bm{\varOmega})
     = \kappa_1 \mathrm{tr}(\bm{\varOmega }^{-1}) - \kappa_2 \mathrm{tr}(\bm{\varLambda}^{-1}\bm{\varPi}^{-2} )
     \\- \kappa_3 \mathrm{tr}(\bm{\varOmega }^{-1} ) \mathrm{tr}(\bm{\varLambda}^{-1}\bm{\varPi}^{-1}) ,
\end{multline}
where $\bm{\varOmega } = \bm{W}\bm{W}^H$, $\bm{\varPi} = \bm{Q}\bm{\varOmega }\bm{Q}^H$, with $\bm{Q}$ containing the eigenvectors of $\bm{R}_{H}$, and $\kappa_1 = \sigma_s^2 N_R/(L-N_T)$, $\kappa_2 = \kappa_1^2$ and $\kappa_3 = \kappa_1^2/(L-N_T)$ are three constants. 

\begin{figure}[!t]
    \centering
    \includegraphics[width=0.4\textwidth]{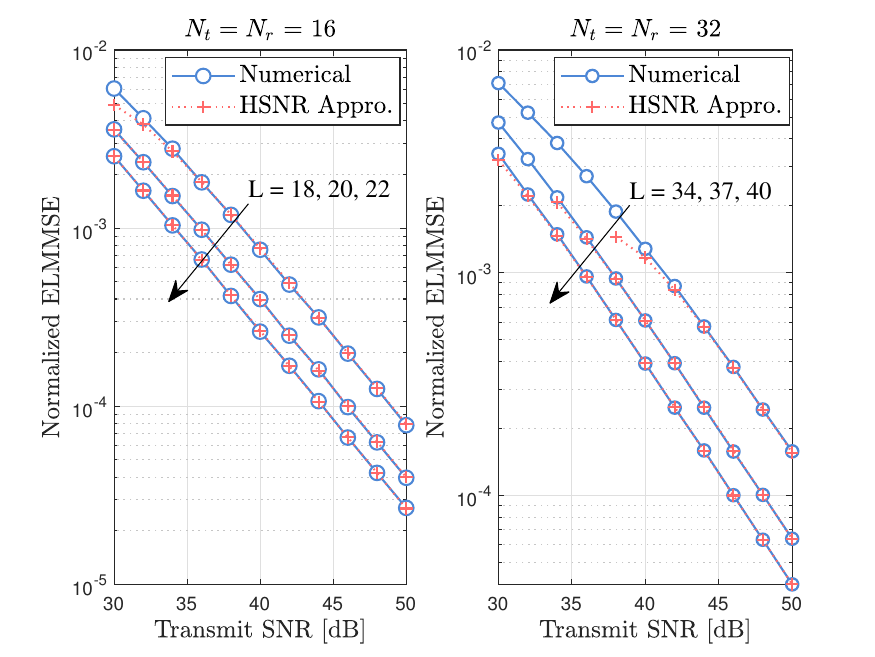}
    \caption{Normalized ELMMSE and its asymptotic formulation \eqref{ELMMSE_HSNR_Appro} versus transmit SNR.} 
    \label{ELMMSE_High_SNR_Appro}
\end{figure}
%\vspace{-1em}

\textbf{{\emph{Proof}}:}
Please refer to Appendix \ref{Lemma_2}. \hfill $\blacksquare$    

As shown in Fig. \ref{ELMMSE_High_SNR_Appro}, we demonstrate the effectiveness of the asymptotic formulation in \eqref{ELMMSE_HSNR_Appro}. It is observed that in the high-SNR regime, the asymptotic formulation of $\mathbb{E}_{\bm{S}}[f(\bm{W} ; \bm{S})]$ fits its theoretical value very well. In what follows, we elaborate on how to leverage the asymptotic formulation to simplify the precoding design for random ISAC signals. Towards that end, we first recast problem \eqref{General_Model_Rand} as
\begin{align}\label{LMMSE_ApproxProblem}
		\min_{\bm{\varOmega }} ~ \tilde{f}(\bm{\varOmega }) 
		~~\mathrm{s.t.} ~ \tr(\bm{\varOmega }) \le P, ~\bm{\varOmega } \succeq \bm{0}.
\end{align}
While problem \eqref{LMMSE_ApproxProblem} is non-convex, its optimal solution has an eigenvalue decomposition structure given in Theorem 2. To show this, let us first introduce a useful lemma below.

% \noindent
\textbf{Lemma 3.} Suppose $\bm{A}$ and $\bm{B}$ are $K \times K $ positive semidefinite Hermitian matrices. Denote their eigen-decompositions by $\bm{A} = \bm{U}_A \bm{\Sigma}_A^{\downarrow} \bm{U}_A^H $ and $\bm{B} = \bm{U}_B \bm{\Sigma}_B^{\downarrow} \bm{U}_B^H $, respectively, where $\bm{\Sigma}_A^{\downarrow} = \mathrm{diag}([\alpha_1,\alpha_2,\dots,\alpha_{K} ])$, $\bm{\Sigma}_B^{\downarrow} = \mathrm{diag}([\beta_1,\beta_2,\dots,\beta_{K}])$, $\alpha_1 \ge \alpha_2 \ge \dots \ge \alpha_K$ and $\beta_1 \ge \beta_2 \ge \dots \ge \beta_K$. Then $\tr(\bm{AB}) \le \sum_{k=1}^{K} \alpha_k \beta_k$ and the upper bound is attained if and only if $\bm{U}_A = \bm{U}_B$.

\textbf{{\emph{Proof}}:}
    Please refer to \cite[H.1.g.]{marshall1979inequalities}.
\hfill $\blacksquare$

% \noindent
\textbf{Theorem 2.} The optimal solution of problem \eqref{LMMSE_ApproxProblem} is
\begin{align}
    \bm{\varOmega }^{\mathsf{opt}} = \bm{Q}^H \bm{\varPsi}^{\uparrow} \bm{Q},
\end{align}
where $\bm{\varPsi}^{\uparrow} = \mathrm{diag}([\psi_1,\psi_2,\dots,\psi_{N_T} ])$ denotes the eigenvalue matrix of $\bm{\varOmega }^{\mathsf{opt}}$ and $\mathrm{diag}(\cdot)$ denotes the diagonal matrix operator. Here, the diagonal elements of $\bm{\varPsi}^{\uparrow}$ are sorted in an increasing
order, i.e., $\psi_1 \le \psi_2 \le\dots \le \psi_{N_T}$.

\textbf{{\emph{Proof}}:}
    Please refer to Appendix \ref{Theorem_2}. \hfill $\blacksquare$ 
    
Theorem 2 suggests that, to minimize the ELMMSE, the optimal DIP precoder should align to the eigenspace of $\bm{R}_H$ in the high-SNR regime. This further simplifies the ELMMSE minimization problem as we only need to numerically compute the eigenvalues in $\bm{\varPsi}$, which is equivalent to allocating the transmit power within the eigenspace of $\bm{R}_H$. For notational convenience, we represent the optimization variables by the vector $\bm{p} = [p_1,p_2,\dots,p_{N_T}]^{T} \in \mathbb{R}^{N_T \times 1}$ with each element satisfying 
\begin{align}\label{Eig_Element}
p_{\ell} = 1/\psi_{N_T+1-\ell},~\ell = 1, 2, \dots, N_T.    
\end{align}
Based on Theorem 2, we recast problem \eqref{LMMSE_ApproxProblem} as
\begin{align}\label{LMMSE_EigValue}
		\min_{\bm{p}} &~~ g(\bm{p}) \triangleq \kappa_1 \sum_{\ell=1}^{N_T} p_{\ell}  - \kappa_2 \sum_{\ell=1}^{N_T} \lambda_{\ell} p_{\ell}^2    - \kappa_3 \sum_{\ell=1}^{N_T} p_{\ell} \sum_{\ell=1}^{N_T} \lambda_{\ell} p_{\ell}   \nonumber\\
		\mathrm{s.t.} &~~ \sum_{\ell=1}^{N_T} \frac{1}{p_{\ell}} \le P, ~p_{\ell} \ge 0,~ \ell = 1,2,\dots,N_T.
\end{align}

Let us denote the feasible region of problem \eqref{LMMSE_EigValue} by $\mathcal{Q}$, which is convex in general. Nevertheless, problem \eqref{LMMSE_EigValue} is still non-convex since the objective function $g(\bm{p})$ is non-convex. We therefore resort to the SCA algorithm to attain a local minimum \cite{liu_Yafeng2022joint}. Let us first approximate $g(\bm{p})$ by using its first-order Taylor expansion at a given feasible point $\bm{p}^{\prime} \in \mathcal{Q}$, expressed as
\begin{align}
    g(\bm{p}) \approx g(\bm{p}^{\prime}) + \langle  \nabla{g}(\bm{p}^{\prime}), \bm{p} - \bm{p}^{\prime} \rangle,
\end{align}
where $\langle\bm{a},\bm{b} \rangle$ represents the inner product of $\bm{a}$ and $\bm{b}$, and $\nabla{g}(\cdot)$ stands for the gradient of $g(\cdot)$ with its $n$-th entry expressed as
\begin{multline}\label{power_allocation_SCA}
        [\nabla{g}(\bm{p})]_{n}    = \kappa_1 - 2 \kappa_2 \lambda_n p_n \\
    - \kappa_3\Big(\sum_{\ell=1}^{N_T} \lambda_{\ell} p_{\ell}+\lambda_n \sum_{\ell=1}^{N_T} p_{\ell}\Big),  n = 1,2, \dots, N_T. 
\end{multline}
At the $j$-th iteration of the SCA algorithm, we solve the following convex optimization problem
\begin{align}\label{LMMSE_EigValue_SCA}
		\min_{\bm{p}} ~ \underline{g}(\bm{p}) \triangleq \langle  \nabla{g}(\bm{p}^{\prime}), \bm{p} - \bm{p}^{\prime} \rangle ~~
		\mathrm{s.t.} ~ \bm{p} \in \mathcal{Q}
\end{align}
to obtain a solution $\bm{p}^{\prime} \in \mathcal{Q}$. Note that $\underline{g}(\bm{p}^{\prime}) \le 0$, since $\underline{g}(\bm{p}^{(j)}) = 0$. This indicates that $\bm{p}^{\prime} - \bm{p}^{(j)}$ yields a descent direction for the $(j+1)$-th iteration. Therefore, one may update the $(j+1)$-th iteration by moving along the descent direction with a certain stepsize $\delta^{(j)}$, which is
\begin{align}
    \bm{p}^{{(j+1)}} = \bm{p}^{(j)} + \delta^{(j)}(\bm{p}^{\prime}-\bm{p}^{(j)}), ~\delta^{(j)} \in [0,1], 
\end{align}
where $\delta^{(j)} \in [0,1]$ ensures $\bm{p}^{(j+1)} \in \mathcal{Q}$ since both $\bm{p}^{(j)}$ and $\bm{p}^{\prime}$ belong to the convex set $\mathcal{Q}$. We summarize the above data-independent SCA algorithm for solving problem \eqref{LMMSE_EigValue} in Algorithm \ref{SCAalg_HighSNR}. By iterating $\bm{p}$ along the descent direction, the above analysis indicates that the objective value of problem \eqref{power_allocation_SCA} is non-increasing after each iteration of Algorithm \ref{SCAalg_HighSNR}. Moreover, the objective value of problem \eqref{power_allocation_SCA} is lower bounded by a finite value due to the limited power budget. Therefore, the proposed SCA algorithm is guaranteed to converge. Denote the output of Algorithm \ref{SCAalg_HighSNR} by $\bm{p}^{\star}$. We may then construct a solution $\bm{\varOmega }^{\star}$ using \eqref{Eig_Element} and Theorem 2.

Now we analyze the complexity of Algorithm \ref{SCAalg_HighSNR}. It requires a complexity $\mathcal{O}(N_T^{3}(2N_T)^{0.5})$ at each iteration \cite{boyd2004convex}. Therefore, the proposed SCA algorithm requires a complexity of $\mathcal{O}(\mathcal{I}_2 N_T^{3.5})$, with $\mathcal{I}_2$ being the maximum allowable number of iterations.

\begin{algorithm}[!t]
    \caption{SCA Algorithm for Solving Problem \eqref{LMMSE_EigValue}.}
    \label{SCAalg_HighSNR}
    \begin{algorithmic}[1]
    \Require
    $P, N_T, N_R, \sigma_s^2, L, \{\kappa_{\ell}\}_{\ell=1}^3, \{\lambda_{\ell}\}_{\ell=1}^{N_T}, j_{\mathrm{max}}, \tau < 0 $.
    \Ensure
    $\bm{p}^{\star}$.
    \State Initialize $j = 1$ and $\bm{p}^{(1)}$ by equal-power allocation.
    \Repeat
    \State Solve the convex problem \eqref{LMMSE_EigValue_SCA} to obatin $\bm{p}^{\prime} \in \mathcal{Q}$ .
    \State Update $\bm{p}^{(j+1)} = \bm{p}^{(j)} + \delta^{(j)}(\bm{p}^{\prime}-\bm{p}^{(j)})$, where $\delta^{(j)}$ can be computed by the exact line search.
    \State Update $j=j+1$.
    \Until $\underline{g}(\bm{p}) \ge \tau $ or $j=j_{\mathrm{max}}$.
    \end{algorithmic}
\end{algorithm}

% \subsection{Complexity Analysis}
\section{ISAC Precoding with Gaussian Signals}\label{Sec_ISAC}

In this section, we focus on the ISAC precoding design to minimize the ELMMSE while satisfying communication performance requirements, which is formulated as the optimization problem \eqref{General_Model_ISAC}. We notice that problem \eqref{General_Model_ISAC} presents a challenging stochastic optimization problem due to the expectation in the objective function as well as the non-convex feasible region. Inspired by the sensing-only precoding design discussed in Sec. \ref{Sec_III}, we first introduce a DDP design for ISAC transmission, which may be solved via a specifically tailored penalty-based AO algorithm. Subsequently, we elaborate on its DIP counterpart, and develop an SGP-AO algorithm to solve the optimization problem. Finally, we leverage the asymptotic formulation of the objective function in the high-SNR regime to further simplify the DIP design.

\subsection{DDP for ISAC: A Penalty-Based AO Method}

Given a realization $\bm{S}_n$, the DDP matrix $\bm{W}_n$ is treated as a function of $\bm{S}_n$. This may introduce additional uncertainty (and thereby extra communication degrees of freedom) for data transmission, in which case the rate achieved in \eqref{GaussiaN_Rate} may no longer be valid since $\bm{W}_n\bm{S}_n$ is non-Gaussian. Therefore, the non-Gaussian $\bm{W}_n\bm{S}_n$ results in the absence of closed-form expression of the channel capacity, which makes it challenging to characterize the communication performance in such a case. Fortunately, the following Proposition 2 indicates that the rate \eqref{GaussiaN_Rate} is still achievable despite that it may not characterize the communication capacity.

% \noindent
\textbf{Proposition 2.}
Suppose that the precoding matrix $\bm{W}$ is dependent on the realization of $\bm{S}$. Then it holds that
\begin{align}
   \mathbb{E}_{\bm{W}}\left[\log \det \left(  \bm{I}_{N_u} + \sigma_c^{-2} \bm{H}_{c} \bm{W} \bm{W}^{H}\bm{H}_{c}^{H} \right)\right] \nonumber \le C,
\end{align}
where $C \triangleq \max I(\bm{Y}_c; \bm{WS} | \bm{H}_c)$ is the capacity of the communication channel under Gaussian signaling and the formula in the left-hand side is achievable.

\textbf{{\emph{Proof:}}}
    Please refer to Appendix \ref{Proposition_2}.
\hfill $\blacksquare$

Following Proposition 2 and by recalling $\bm{W}_n \triangleq \bm{W}(\bm{S}_n)$, one may observe that if 
\begin{align}
    {R}(\bm{W}_n) \triangleq \log \det \left(  \bm{I}_{N_u} + \sigma_c^{-2} \bm{H}_{c} \bm{W}_n \bm{W}_n^{H}\bm{H}_{c}^{H} \right) \ge R_0, \forall n,\nonumber
\end{align}
then 
\begin{align}
    C \ge \;& \mathbb{E}_{\bm{W}}\left[\log \det \left(  \bm{I}_{N_u} + \sigma_c^{-2} \bm{H}_{c} \bm{W} \bm{W}^{H}\bm{H}_{c}^{H} \right)\right] \nonumber\\
    = \;& \lim_{N\to\infty}\frac{1}{N} \sum_{n=0}^{N-1} {R}(\bm{W}_n)  
    \ge  R_0. \nonumber
\end{align}
Accordingly, we restrict the constraint as ${R}(\bm{W}_n) \ge R_0$ for all $n = 1, 2, \dots, N$, to facilitate the decomposition of the achievable rate constraint. In line with the sensing-only DDP design presented in Sec. \ref{Sec_III}, the ISAC DDP design aims to solve the following $N$ parallel sub-problems:
% \begin{align}\label{Data_Dep_ISAC}
% 		\min_{\{ \bm{W}(\bm{S}_n)\}_{n=1}^{N}} &~~ \frac{1}{N} \sum_{n=0}^{N-1} {f}(\bm{W}(\bm{S}_n); \bm{S}_n) \nonumber \\ 
% 		 ~~\mathrm{s.t.} &~~R(\bm{W}(\bm{S}_n)) \ge R_0, \bm{W}(\bm{S}_n ) \in \mathcal{A}.
% \end{align}  
% Let $\bm{W}_n \triangleq \bm{W}(\bm{S}_n)$ for notional convenience. Since different $\bm{S}_n$ is independent of each other, we can decompose problem \eqref{Data_Dep_ISAC} as 
\begin{align}\label{ELMMSE_Rate_op}
    \min_{\bm{W}_n \in \mathcal{A}} ~ {f}(\bm{W}_n; \bm{S}_n)  ~~
		\mathrm{s.t.} ~ {R}(\bm{W}_n) \ge R_0.
\end{align}
Again, problem \eqref{ELMMSE_Rate_op} is non-convex and is thus difficult to solve directly. We first introduce an auxiliary variable $\bm{\varOmega }_{n} = \bm{W}_{n}\bm{W}_{n}^H$. Then the rate constraint is recast as $\tilde{R}(\bm{\varOmega }_{n}) \triangleq \log \det \left(  \bm{I}_{N_u} + \sigma_c^{-2} \bm{H}_{c}\bm{\varOmega }_{n}\bm{H}_{c}^{H} \right) \ge R_0 $. Therefore, each sub-problem \eqref{ELMMSE_Rate_op} may be re-expressed as
\begin{align}\label{ELMMSE_Rate_SGP_Op1}
    \min_{\bm{W}_n \in\mathcal{A},\bm{\varOmega }_{n}} &~ {f}(\bm{W}_n; \bm{S}_n)  ~~ \nonumber \\
		\mathrm{s.t.}~~~ &~ \tilde{R}(\bm{\varOmega }_{n}) \ge R_0, \quad \bm{\varOmega }_{n} = \bm{W}_{n}\bm{W}_{n}^H. 
\end{align}
To proceed, we propose penalizing the equality constraint $\bm{\varOmega }_{n} = \bm{W}_{n}\bm{W}_{n}^H$ into the objective function, yielding
\begin{align}\label{DDP_subproblem}
     \min_{\bm{W}_n \in\mathcal{A},\bm{\varOmega }_{n}} &~ {f}(\bm{W}_n; \bm{S}_n) + \frac{\rho}{2} \| \bm{\varOmega }_{n} - \bm{W}_{n}\bm{W}_{n}^H \|_F^2
    \nonumber\\
		\mathrm{s.t.}~~~ &~\tilde{R}(\bm{\varOmega }_{n}) \ge R_0,
\end{align}
where $\rho > 0$ is the penalty parameter. Notice that this problem is convex with respect to the auxiliary variable $\bm{\varOmega }_{n}$ but non-convex on the precoding matrix $\bm{W}_n$. We thus propose to solve the problem by the AO algorithm, as detailed below.

Specifically, with given $\bm{W}_n^{(t)}$ at the $t$-th iteration, we solve the following optimization subproblem:
\begin{align}\label{iter_Xn}
     \min_{\bm{\varOmega }_{n}} ~~ \frac{\rho}{2} \| \bm{\varOmega }_{n} - \bm{W}_{n}^{(t)}(\bm{W}_{n}^{(t)})^H \|_F^2  ~~
		\mathrm{s.t.} ~ \tilde{R}(\bm{\varOmega }_{n}) \ge R_0.
\end{align}
Observe that this problem is convex, which can be solved by off-the-shelf numerical toolbox such as CVX \cite{boyd2004convex}. Denote the solution by $\bm{\varOmega }_{n}^{(t)}$ at the $t$-th iteration. To update $\bm{W}_n$, we need to solve the following constrained non-convex subproblem:
\begin{align}\label{GD_Update_W}
     \min_{\bm{W}_n \in\mathcal{A}} {h}(\bm{W}_n)\triangleq {f}(\bm{W}_n ; \bm{S}_n) + \frac{\rho}{2} \| \bm{\varOmega }_{n}^{(t)} - \bm{W}_{n}\bm{W}_{n}^H \|_F^2.
\end{align}
Despite the non-convex objective function, the feasible region of \eqref{GD_Update_W} is convex, which admits a computable projection step to ensure feasibility. To proceed with the gradient projection algorithm, let us first compute the gradient of $h(\bm{W}_n)$ as
\begin{align}
    \nabla{h}(\bm{W}_n) = \nabla{f}(\bm{W}_n;\bm{S}_n)+ \rho \big(\bm{W}_n\bm{W}_n^H -\bm{\varOmega }_{n}^{(t)}\big) \bm{W}_n,
\end{align}
in which $\nabla {f}(\bm{W}_n;\bm{S}_n)$ is calculated by \eqref{SGD_Gradient}. Therefore, we can update the precoding matrix by 
\begin{align}\label{update_Wn}
    \bm{W}_n^{(t+1)}  \leftarrow \mathsf{Proj}_{\mathcal{A}}\big(\bm{W}_n^{(t)} - \nu ^{(t)} \nabla{h}(\bm{W}_n^{(t)})\big), 
\end{align}
where $\nu ^{(t)}$ denotes the step size.

\begin{algorithm}[!t]
    \caption{Penalty-Based AO Algorithm for Solving \eqref{ELMMSE_Rate_op}.}
    \label{AOalg_DataDep}
    \begin{algorithmic}[1]
    \Require
    $\bm{R}_H, \bm{H}_c, R_0, P, N_T, N_R, \sigma_s^2, \sigma_c^2, L, t_{\mathrm{max}}, \tau_0 ,\xi_0$.
    \Ensure
    $\bm{W}^{\star}_n = \bm{W}_n^{(t)}, n = 1, 2, \dots, N$.
    \State Initialize the Gaussian signal set $\mathcal{S} = \{\bm{S}_1, \bm{S}_2, \ldots, \bm{S}_N\}$.
    % \State Initialize $t = 1$, $\rho = 1$, $\nu^{(t)} = 1$.
    \State Initialize $t = 1$ and $\bm{W}_n^{(1)}$ by water-filling on $\bm{H}_c$.
    \Repeat
    \State Solve the problem \eqref{iter_Xn} to obtain $\bm{\varOmega }_n^{(t)}$ by CVX.
    % \State Increase penalty by $\rho = 1.1 \rho$.
    % \State Decrease step size by $\nu^{(t)} = \frac{1}{(1+t)^{0.8}}$ 
    \State Update the precoding matrix $\bm{W}_n^{(t)}$ by \eqref{update_Wn}.
    \State Update $t=t+1$.
    \Until the decrease of the objective value is below $\tau_0$ and $\xi \le \xi_0$, or $t=t_{\mathrm{max}}$.
    \end{algorithmic}
\end{algorithm}

Now we are ready to present the penalty-based AO algorithm for the DDP design in ISAC scenarios, as detailed in Algorithm \ref{AOalg_DataDep}. By iteratively updating the variables, it is observed that the objective function in \eqref{ELMMSE_Rate_SGP_Op1} remains monotonically non-increasing from each $t-$th iteration to $(t+1)-$th iteration. Moreover, the objective function in \eqref{ELMMSE_Rate_SGP_Op1} is bounded due to the limited transmit power budget. With any fixed $\rho$, it is observed that the achieved objective value of problem \eqref{ELMMSE_Rate_SGP_Op1} is an upper bound of that achieved by the original problem \eqref{ELMMSE_Rate_op}. By alternatively solving the problem \eqref{iter_Xn} and \eqref{GD_Update_W}, this upper bound can be gradually tightened. It is noted that the penalty parameter $\rho$ is increased in each iteration to force the equality constraint $\| \bm{\varOmega }_{n} - \bm{W}_{n}\bm{W}_{n}^H \|_F^2$ to be ultimately satisfied. As such, a stationary point is attained in Algorithm \ref{AOalg_DataDep}. Note that the performance of Algorithm \ref{AOalg_DataDep} relies on the choice of $\rho$ since the converged solution may not satisfy the communication performance requirement. To tackle this issue, we examine the following violation indicator
\begin{align}
    \xi = | R(\bm{W}_n^{(t)}) - R_0 | \le \xi_0,
\end{align}
for terminating the iteration, where $\xi_0$ is a tolerable threshold. In general, a larger value of $\rho$ may achieve better performance but at the cost of more iterations. To this end, we gradually increase the value of the penalty factor $\rho$ and decrease the step size $\nu^{(t)}$ to promote convergence. Moreover, one may use the well-known water-filling solution of rate maximization as a good initial point.

Finally, we analyze the complexity of Algorithm \ref{AOalg_DataDep}. By solving problem \eqref{iter_Xn} and \eqref{GD_Update_W}, the complexities of updating $\bm{\varOmega }_{n}$ and $\bm{W}_n$ and are of order $\mathcal{O}(N_T^{7})$ and $\mathcal{O}(N_T^3+2N_T^2)$, respectively. Therefore, the proposed penalty-based AO algorithm requires a complexity at the order of $\mathcal{O}\big( \mathcal{I}_3 N_T^{2}(N_T^{5} + N_T + 2 )\big)$, where $\mathcal{I}_3$ represents the number of iterations required.

\subsection{DIP for ISAC: A Penalty-Based SGP-AO Method}
We then apply the AO framework for the DIP design in ISAC scenarios. Following a similar procedure above, let us first introduce an auxiliary variable $\bm{\varOmega } = \bm{W}\bm{W}^H$, which is the covariance matrix of transmitted signals. In such a case, the communication channel capacity is characterized by
\begin{align}
    \tilde{R}(\bm{\varOmega }) = \log \det \left(  \bm{I}_{N_u} + \sigma_c^{-2} \bm{H}_{c}\bm{\varOmega }\bm{H}_{c}^{H} \right),
\end{align}
since $\bm{W}$ is independent of $\bm{S}$. By penalizing $\bm{\varOmega } = \bm{W}\bm{W}^H$ into the objective function, problem \eqref{General_Model_ISAC} is recast as
\vspace{-0.5em}
\begin{align}\label{ELMMSE_Rate_SGP_Op}
     \min_{\bm{W} \in\mathcal{A},\bm{\varOmega }} &~ \mathbb{E}_{\bm{S}}[f(\bm{W} ; \bm{S})] + \frac{\rho }{2} \| \bm{\varOmega } - \bm{W}\bm{W}^H \|_F^2 ~~ \nonumber \\
		\mathrm{s.t.} ~~&~ \tilde{R}(\bm{\varOmega }) \ge R_0.
\end{align}
It is worth pointing out that problem \eqref{ELMMSE_Rate_SGP_Op} is a stochastic version of problem \eqref{DDP_subproblem}. The key difference lies in the utilization of $\hat{\nabla} {f}(\bm{W}) + \rho  (\bm{W}\bm{W}^H -\bm{\varOmega }) \bm{W}$ in \eqref{update_Wn} as the stochastic gradient to update $\bm{W}$. Then $\bm{W}$ and $\bm{\varOmega }$ can be updated in an alternating manner, which is referred to as the SGP-AO algorithm. Since the main procedure of SGP-AO closely resembles Algorithm \ref{AOalg_DataDep}, we omit further details here for brevity.

\begin{figure*}[!t]
\centering
\subfigure[Batch size $|\mathcal{D}^{(r)}| = 1$.] { \label{MBSGP_SGP_bitchsize1_Iteration_vsSNR}
			\includegraphics[width=0.31\textwidth]{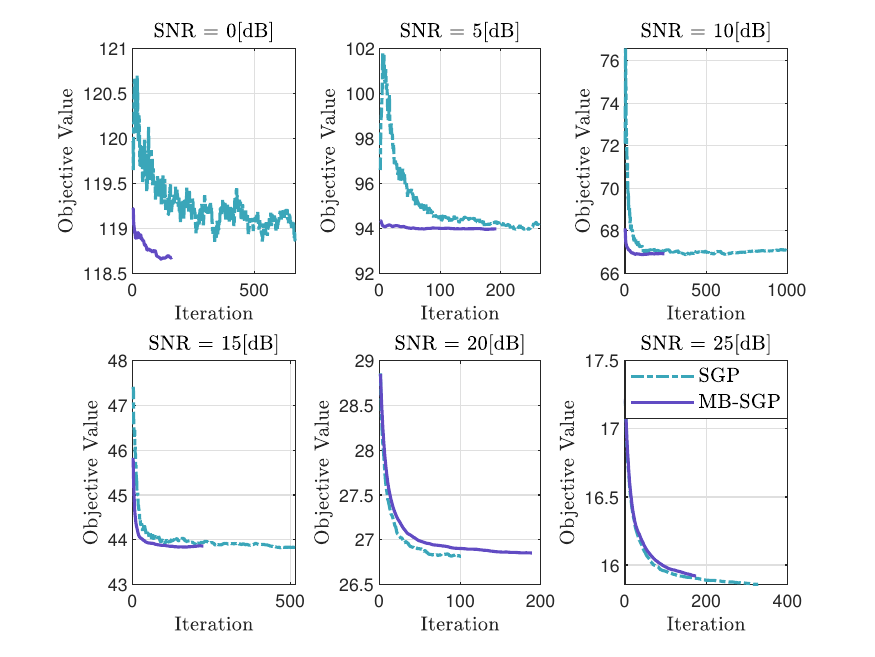}
		}
\subfigure[Batch size $|\mathcal{D}^{(r)}| = 10$.] { \label{MBSGP_SGP_bitchsize1_error_vsSNR}
			\includegraphics[width=0.31\textwidth]{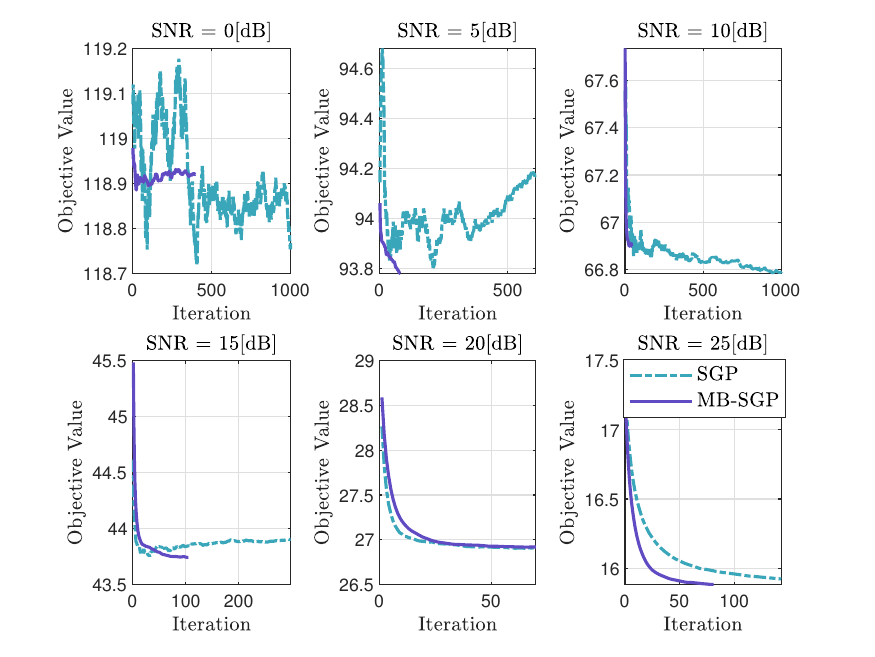}
		}
\subfigure[Batch size $|\mathcal{D}^{(r)}| = 100$.] { \label{MBSGP_SGP_bitchsize100_error_vsSNR}
			\includegraphics[width=0.31\textwidth]{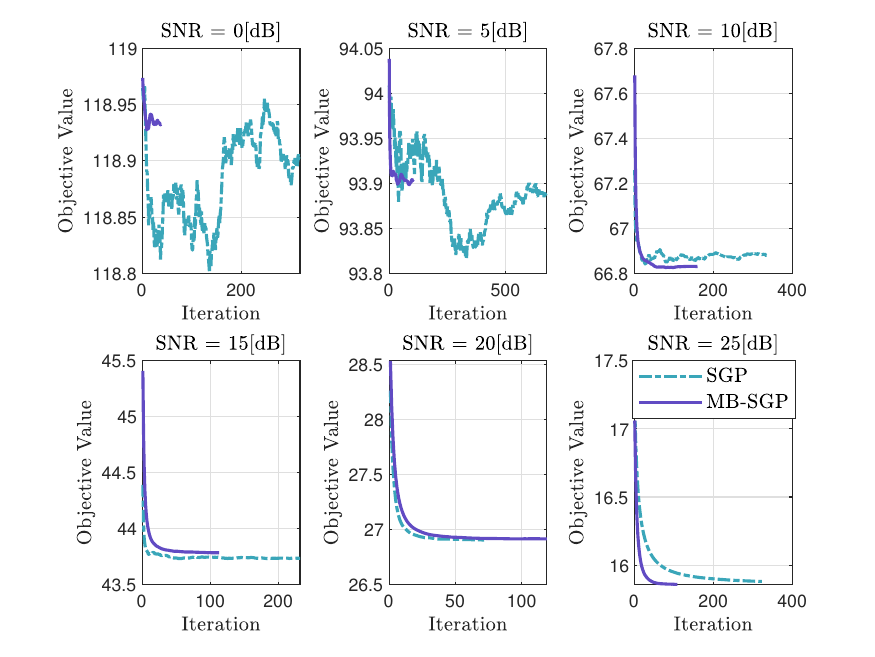}
		}
\caption{The convergence performance of the proposed MB-SGP algorithm and the proposed SGP algorithm versus different SNR settings.}
        \label{MBSGP_SGP_convergence}
\end{figure*} 

\subsection{DIP for ISAC in the High-SNR Regime}
We observe that the main challenge of solving problem \eqref{General_Model_ISAC} lies in the fact that the feasible region is non-convex and the objective function $\mathbb{E}_{\bm{S}}[f(\bm{W}; \bm{S})]$ is in an expectation form. Fortunately, the feasible region can be recast as a convex set with respect to $\bm{\varOmega } = \bm{WW}^H$, i.e.,
\begin{align}
    \mathcal{A}_{\varOmega} = \big\{ \bm{\varOmega }  & \in  ~\mathbb{C}^{N_T \times N_T} ~|~ \mathrm{tr}(\bm{\varOmega } ) \le P, \nonumber \\
    &\log \det \left(  \bm{I}_{N_u} + \sigma_c^{-2} \bm{H}_{c}\bm{\varOmega } \bm{H}_{c}^{H} \right) \ge R_0 \big\}.
\end{align}
Inspired by the asymptotic formulation in the high-SNR regime, we first approximate $\mathbb{E}_{\bm{S}}[f(\bm{W}; \bm{S})]$ by $ \tilde{f}(\bm{\varOmega })$ in Proposition 1. Therefore, the stochastic optimization problem \eqref{General_Model_ISAC} is recast as a deterministic optimization problem, which is denoted by 
$\min_{\bm{\varOmega } \in \mathcal{A}_{\varOmega}}   \tilde{f}(\bm{\varOmega })$.
% \begin{align}\label{ELMMSE_Rate_HSNR}
% 		\min_{\bm{\varOmega } \in \mathcal{A}_{\varOmega}}   \tilde{f}(\bm{\varOmega }).
% \end{align}

Notice that the objective function $\tilde{f}(\bm{\varOmega })$ is still non-convex. To solve this problem, we propose to harness the established SCA framework in Algorithm \ref{SCAalg_HighSNR}. The key procedure is similar to Algorithm \ref{SCAalg_HighSNR} in Sec. \ref{Sec_III}. Hence, for brevity, we do not delve into the details here and only present the first-order approximation as well as the initialization in the sequel. Given a feasible point $\bm{\varOmega }^{\prime} \in \mathcal{A}_{\varOmega}$, the first-order expansion approximation of $\tilde{f}(\bm{\varOmega })$ is 
\begin{align}
     \tilde{f}(\bm{\varOmega })\approx \tilde{f}(\bm{\varOmega }^{\prime}) + \langle \nabla\tilde{f}(\bm{\varOmega }^{\prime}), \bm{\varOmega }  - \bm{\varOmega }^{\prime}\rangle,
\end{align}
where $\nabla\tilde{f}(\cdot)$ is calculated by
\begin{multline}
     \nabla\tilde{f}(\bm{\varOmega }) = -\kappa_1 \bm{\varOmega }^{-2} + 2 \kappa_2 (\bm{Q}^H \bm{\varLambda}^{-1}\bm{\varPi}^{-3} \bm{Q}) \\ + \kappa_3 \tr(\bm{\varLambda}^{-1}\bm{\varPi}^{-1})\bm{\varOmega }^{-2} + \kappa_3 \tr(\bm{\varOmega }^{-1})\bm{Q}^H\bm{\varLambda}^{-1}\bm{\varPi}^{-2}\bm{Q}. 
\end{multline}
% To proceed SCA algorithm, we omit the constant term $\tilde{f}(\bm{\varOmega }^{i})$ and solve the following convex optimization problem at $(i+1)$-th each iteration:
% \begin{align}\label{HSNR_SCA}
% 		\min_{\bm{\varOmega } \in \mathcal{A}_{\varOmega}} 
%   \underline{f}(\bm{\varOmega }) \triangleq \langle \nabla\tilde{f}(\bm{\varOmega }^{i}), \bm{\varOmega }  - \bm{\varOmega }^{i}\rangle
% \end{align}
% where $\bm{\varOmega }^{i} \in \varOmega $ is obtained at the $i$-th iteration. By solving problem \eqref{HSNR_SCA}, we obtain an optimized solution $\bm{\varOmega }^{\prime}$. At each iteration, $(\bm{\varOmega }^{\prime}-\bm{\varOmega }^{i})$ yields a descent direction since $\tilde{f}(\bm{\varOmega }^{\prime}) \le \tilde{f}(\bm{\varOmega }^{i})$. We are now ready to get the updated point $\bm{\varOmega }^{i+1}$, expressed as
% \begin{align}
%     \bm{\varOmega }^{i+1} = \bm{\varOmega }^{i} + \upsilon ^{i} \left(\bm{\varOmega }^{\prime} - \bm{\varOmega }^{i} \right), ~\upsilon ^{i} \in [0,1],
% \end{align}
% where $\upsilon ^{i}$ denotes the step size. It is straightforward to find that $\bm{\varOmega }^{i+1} \in \varOmega $ since both $\bm{\varOmega }^{i}$ and $\bm{\varOmega }^{\prime}$ are drawn form the convex feasible set $\varOmega $.
A good initial point may be obtained by solving the convex problem
% :
% \begin{align}
%     \min_{\bm{\varOmega } \in \mathcal{A}_{\varOmega}} \tr(\bm{\varOmega }^{-1}).
% \end{align}
$\min_{\bm{\varOmega } \in \mathcal{A}_{\varOmega}} \tr(\bm{\varOmega }^{-1})$.

\section{Simulation Results}\label{Sec_V}

\begin{figure*}[!t]
\centering
  \subfigure[Iterations at $|\mathcal{D}^{(r)}| = 10$.] { \label{MBSGP_SGP_bitchsize10_Iteration_vsSNR}
			\includegraphics[width=0.23\textwidth]{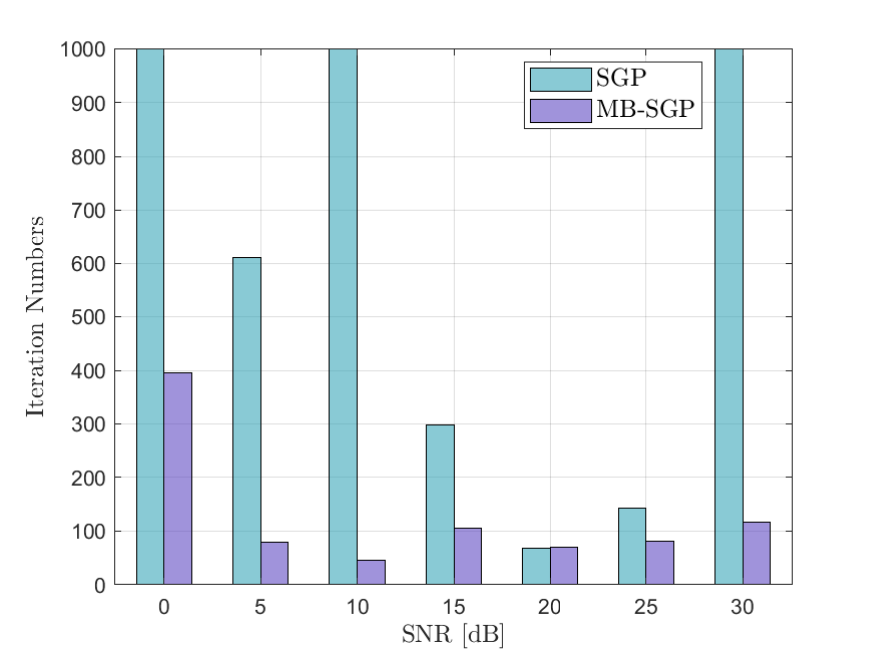}
		}
    \subfigure[Iterations at $|\mathcal{D}^{(r)}| = 100$.] { \label{MBSGP_SGP_bitchsize100_Iteration_vsSNR}
			\includegraphics[width=0.23\textwidth]{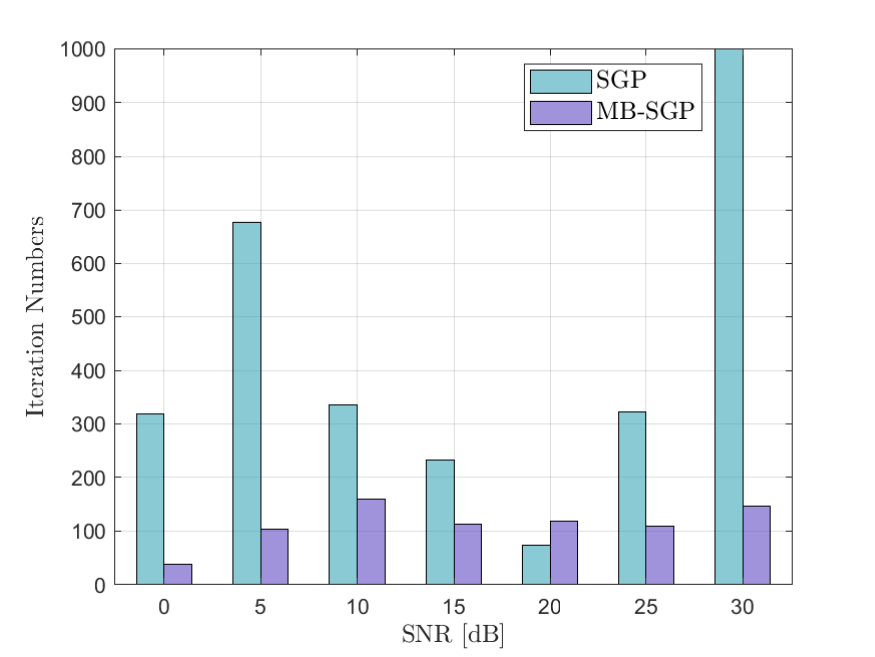}
		}
		\subfigure[ELMMSE at $|\mathcal{D}^{(r)}| = 10$.] { \label{MBSGP_SGP_bitchsize10_error_vsSNR}
			\includegraphics[width=0.23\textwidth]{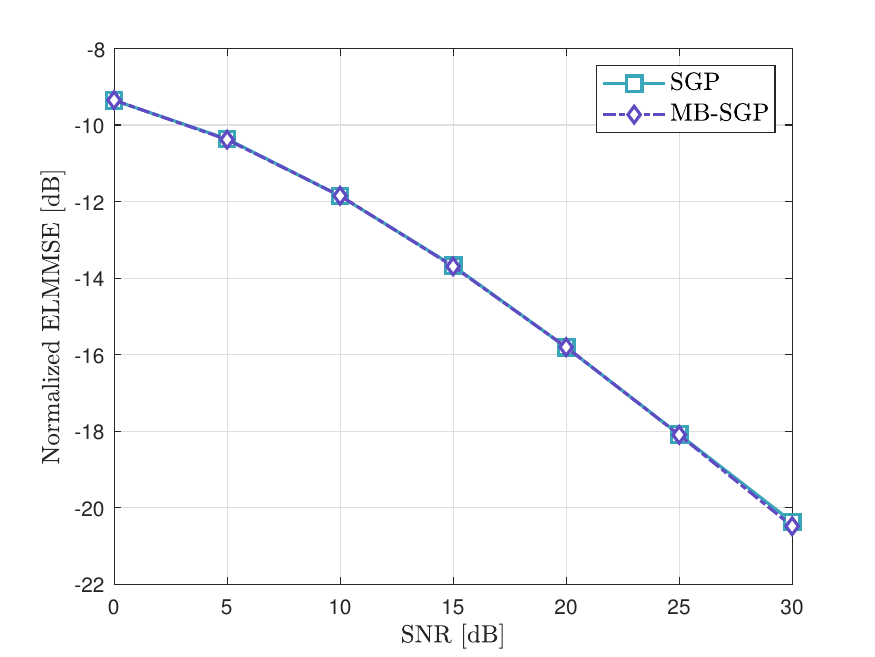}
		}.
  	\subfigure[ELMMSE at $|\mathcal{D}^{(r)}| = 100$.] { \label{MBSGP_SGP_bitchsize100_error_vsSNR}
			\includegraphics[width=0.23\textwidth]{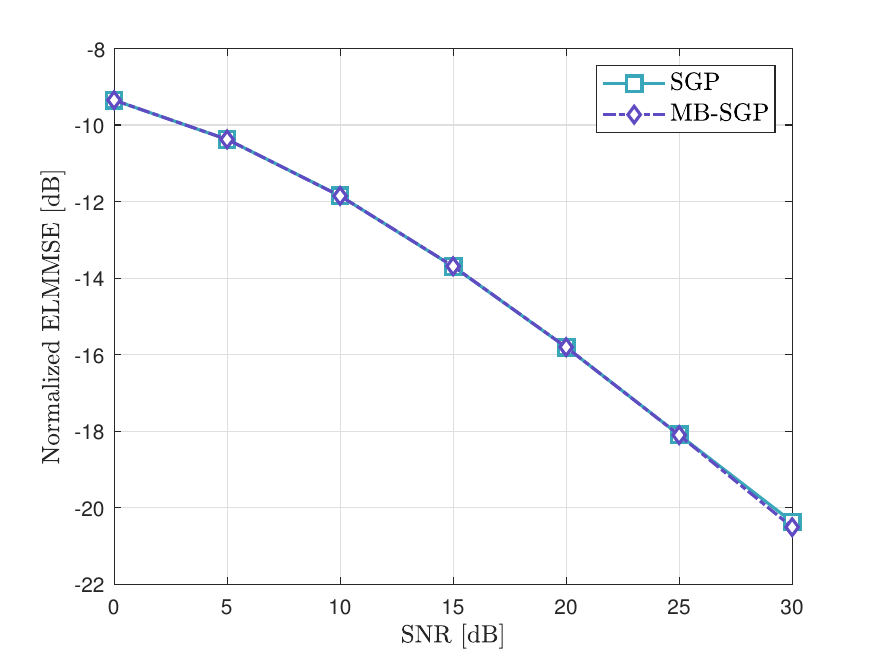}
		}
		\caption{iteration number and normalized ELMMSE performance of using MB-SGP and SGP algorithms versus different SNR settings.}
        \label{MBSGP_SGP_iteration_error}
\end{figure*}

In this section, we demonstrate the numerical results of the proposed methods in both sensing-only and ISAC scenarios. Unless otherwise specified, the transmit SNR is defined as $P/ \sigma_s^2$. The simulation parameters are listed in Table \ref{tab1}. In both SGP and penalty-based SGP-AO methods, we set the number of mini-batch samples as $|\mathcal{D}^{(r)}| = 10$. The eigenvalues of $\bm{R}_H$ represent the spatial channel correlation, which follow a uniform distribution on the interval $[1,10]$ in the simulation.

\begin{table}[!t]    
	\centering
	\fontsize{9.0}{8.0}\selectfont
	\caption{Parameters in Simulations}\label{tab1}
		{\begin{tabular}{p{1.0cm} p{1.0cm} p{1.0cm} p{1.0cm}l p{1.0cm} p{1.0cm}}
				\hline 
				\hline 
				Parameter & Value & Parameter & Value & Parameter & Value \\
				\hline
				$N_T$ & $32$ &	$N_R$ & $32$ & $N_u$ & $4$ \\				
				$\sigma_s^2$	& $0~\mathrm{dBm}$ & $\sigma_s^2$	& $0~\mathrm{dBm}$  &  $N$	& $100$ \\
				$a$ & $10$ & $b$ & $10$ &  $r_{\mathrm{max}}$ & $1000$ 	\\
                $\epsilon$ & $10^{-5}$ &$\xi_0$  & $0.1$ & $\tau$  & $-10^{-5}$ \\
                $j_{\mathrm{max}}$ & $30$ &	$\tau_0$ & $10^{-3}$	& $t_{\mathrm{max}}$  & $30$ \\
                $\beta_1$ & $0.6$  & $\beta_2$  & $0.999$  &$\varepsilon_0$ & $-10^{-8}$\\
				\hline 
		\end{tabular}}
\end{table}

\subsection{Convergence Examples}\label{conv_subsec}

First, let us examine the convergence performance of our proposed algorithms in both sensing-only and ISAC scenarios. As for the sensing-only scenario, we show the convergence of our proposed SGP and MB-SGP algorithms versus different numbers of mini-batch samples $|\mathcal{D}^{(r)}|$ in Fig. \ref{MBSGP_SGP_convergence} and Fig. \ref{MBSGP_SGP_iteration_error}, with fixed $L$ = 32 but different SNR settings.  We set the iteration step size as $\eta^{(r)} = 10/(10+r)$ at the $r$-th iteration, as recommended for ensuring the convergence \cite{li2019convergence}. Notably, both SGP and MB-SGP algorithms can be implemented {\it offline} based on the locally generated Gaussian signal samples. Moreover, both of them exhibit a rapid convergence when $|\mathcal{D}^{(r)}| = 10$, striking a tradeoff between complexity and convergence when compared to $|\mathcal{D}^{(r)}| = 1$ (low complexity but poor convergence) and $|\mathcal{D}^{(r)}| = 100$ (high complexity but favorable convergence).

In Fig. \ref{MBSGP_SGP_iteration_error}, we provide a more detailed comparison of the performance between the developed MB-SGP algorithm and the SGP algorithm. It is observed that under different SNR parameters, the MB-SGP algorithm requires fewer iterations to converge compared to the SGP algorithm. Furthermore, combining with Fig. \ref{MBSGP_SGP_bitchsize10_error_vsSNR}-\ref{MBSGP_SGP_bitchsize100_error_vsSNR}, it is concluded that the estimation performance of the developed  MB-SGP algorithm is comparable to that of the SGP algorithm, while significantly reducing the number of iterations and accelerating algorithm convergence. This demonstrates the superiority of the proposed MB-SGP algorithm over the classical SGP algorithm.

\begin{figure}[t!]
	\centering
	\includegraphics[width=0.35\textwidth]{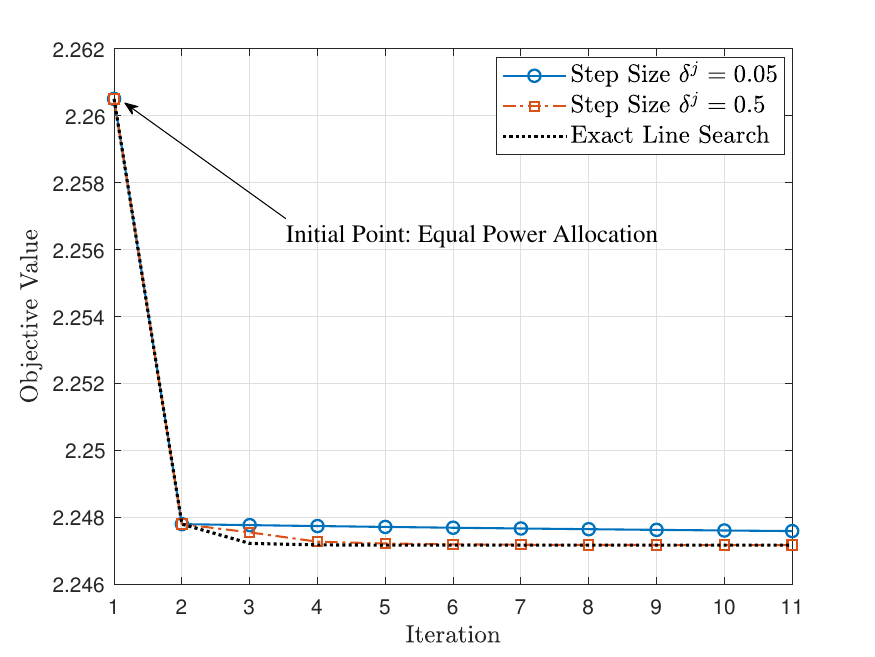}
	\caption{SCA algorithm in the high-SNR regime in sensing-only scenarios.}
        \label{SenOnly_HSNR_SCA}
\end{figure}

In Fig. \ref{SenOnly_HSNR_SCA}, we show the convergence of our proposed SCA algorithm for solving \eqref{LMMSE_EigValue} in the high-SNR regime, where the transmit SNR is fixed to be $32$ dB and the frame length is set as $L = 40$. It is shown that SCA converges quickly within $10$ iterations and the converged objective value is very close to its initial point which is determined by equal power allocation. This result indicates that in the high-SNR region, the SGP-based DIP design shall allocate power equally over the eigenspace of $\bm{R}_H$. Interestingly, this strategy aligns with the performance of the traditional water-filling strategy in the high-SNR regime. This provides us with an intuitive insight that, at a high-SNR regime, both of the traditional water-filling scheme and SGP-based DIP scheme will converge to equal-power allocation. We will illustrate this insight in Fig. \ref{SenOnly_HNSR_Algcompare}.

\begin{figure}[t!]
    \centering
    \includegraphics[width=0.45\textwidth]{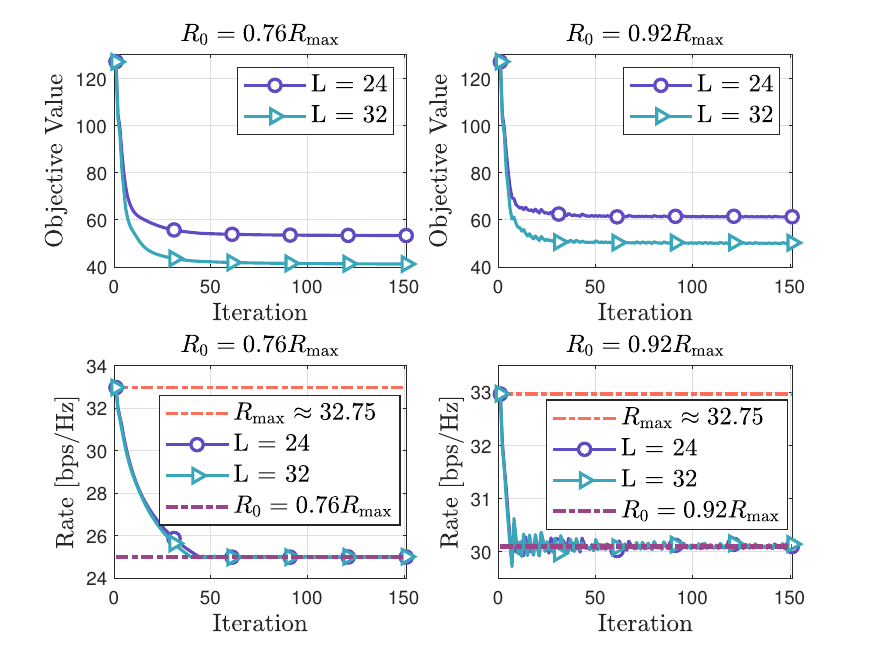}
    \caption{Penalty-based SGP-AO algorithm in ISAC scenarios.}
    \label{ISAC_SGPAO_Convergence}
\end{figure}

In Fig. \ref{ISAC_SGPAO_Convergence}, we show the convergence performance of our proposed penalty-based SGP-AO algorithm in the ISAC scenario. 
%At each iteration $t$, we decrease the step size by setting $\hat{\eta}^{(t)} = 1/(1+t)$ and increase the penalty factor by setting $\varrho  = 0.8(1+t)$. 
The maximum communication rate is obtained by classical water-filling solution over the singular values of $\bm{H}_c$. It can be found that our proposed penalty-based SGP-AO algorithm quickly converges within tens of iterations. Moreover, our proposed algorithm not only exhibits excellent convergence performance but also ensures compliance with penalty constraints and communication rate constraints under different parameter settings.

\subsection{Sensing-Only Precoding}
In this subsection, we aim to evaluate the performance of
our proposed precoding schemes DDP and DIP in Sec. \ref{Sec_III} for sensing-only scenarios. Our benchmark technique is the water-filling scheme proposed in \cite{biguesh2006training}, as given in \eqref{LMMSE_Determinstic_Opt_W}.

\begin{figure*}[!h]
    \centering
    \subfigure[$L : N_T = 3 : 4$.]{
    \includegraphics[width=0.315\textwidth]{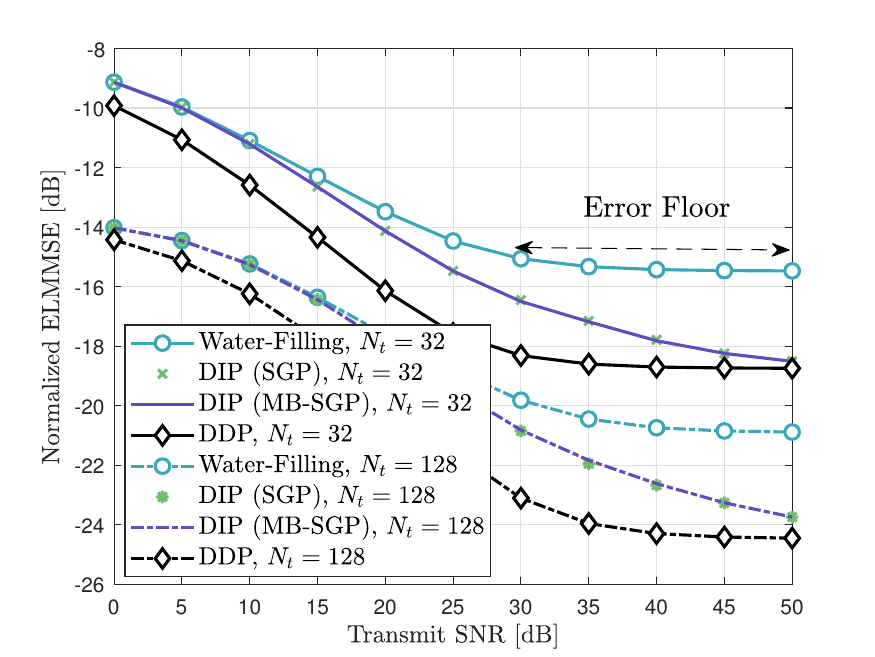} \label{L_small_Nt}
    }
    \subfigure[$L : N_T = 1 : 1$.]{
    \includegraphics[width=0.315\textwidth]{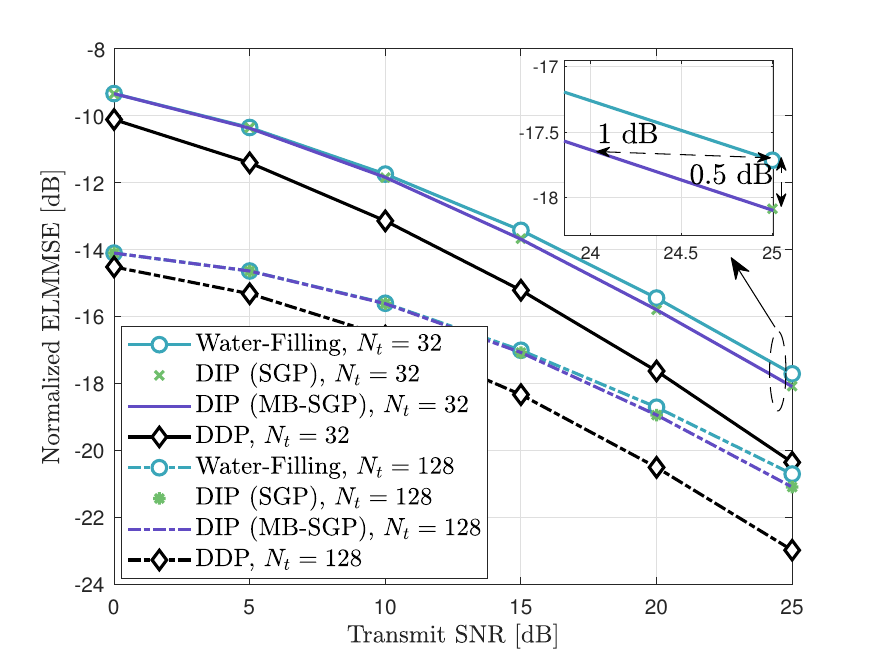}
    \label{L_equal_Nt}
    }
    \label{SenOnly_SGP_WF_DDP_SNR}
    \subfigure[A high-SNR regime.]{
    \includegraphics[width=0.315\textwidth]{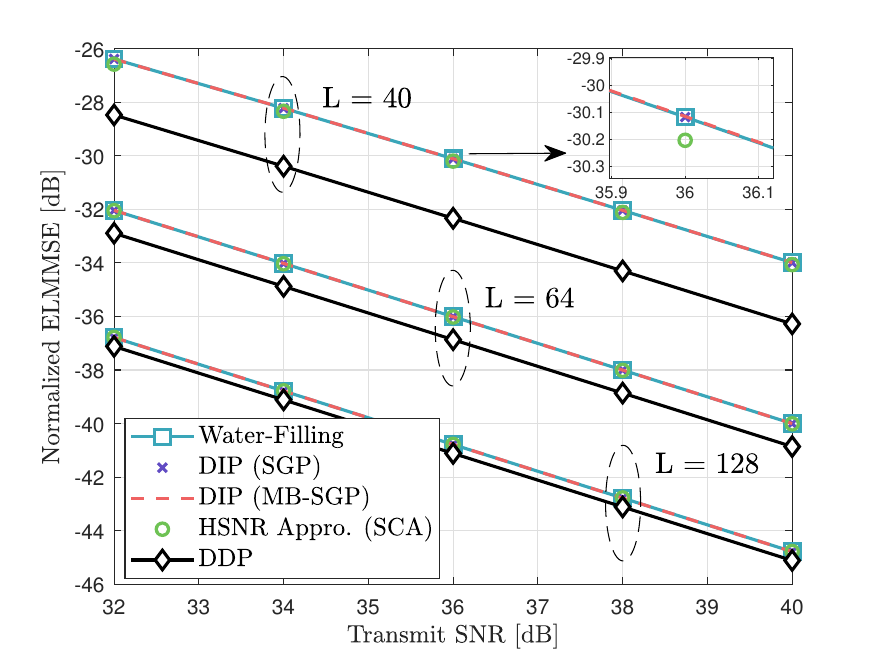}\label{SenOnly_HNSR_Algcompare}
    }\caption{Performance comparison among different precoding schemes versus the transmit SNR in sensing-only scenarios.}
    \label{Sen_Only_Performace} 
\end{figure*} 

In Fig. \ref{L_small_Nt} and Fig. \ref{L_equal_Nt}, we compare three precoding schemes, namely, classical water-filling, DDP, and DIP under different frame length and arry size settings. It is observed that both the number of antennas and frame length significantly impact the sensing performance. Notably, our proposed DDP and DIP schemes outperform the classical water-filling scheme in both cases with $L < N_T$ and $L = N_T$. This can be attributed to two primary factors as follows: Firstly, when $L < N_T$, the rank-deficient nature of the data samples (as shown in Fig. \ref{L_small_Nt}) introduces an error floor in the water-filling scheme, rendering it suboptimal even for deterministic signals. Secondly, the classical water-filling approach overlooks the random nature of the ISAC signals, leading to a performance loss in terms of the ELMMSE for both $L < N_T$ and $L = N_T$. Moreover, it is observed that our proposed DDP scheme maintains the best performance consistently, due to the precise knowledge of transmitted data. However, adapting data-dependent precoding to varying data realizations necessitates intricate modifications, resulting in substantial complexity. Fortunately, the DIP approach can be pre-implemented offline using SGP, striking a favorable tradeoff between performance and complexity.

In Fig. \ref{SenOnly_HNSR_Algcompare}, we show the performance comparison by considering the high-SNR regime. The method ``HSNR Appro. (SCA)'' refers to Algorithm \ref{SCAalg_HighSNR}. It is easily observed that the sensing performance achieved by random ISAC signaling experiences a substantial improvement with increasing SNR and frame length $L$. Moreover, there is a tiny sensing performance gap between ``DIP (SGP)'', ``HSNR Appro. (SCA)'', and ``Water-Filling'', which is consistent with the earlier discussion in Sec. \ref{conv_subsec}. Indeed, all of the above three precoding schemes are independent of data realizations. They exhibit performance close to DDP with an increasing frame length (see $L = 128$) in the high-SNR regime.

%This result indicates that even if the signal is random, the loss in sensing performance due to randomness can be compensated for by accumulating a sufficiently long frame length (compared to the number of antennas) or increasing SNR. 
%\vspace{-0.5em}
\begin{figure}[!t]
	\centering
	\includegraphics[width = 0.43\textwidth]{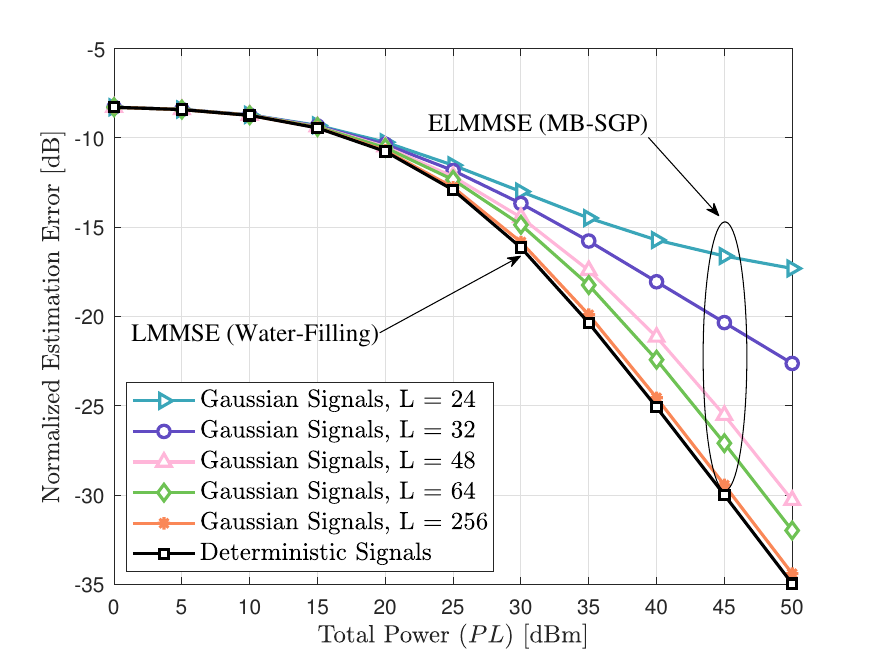}
	\caption{Performance comparison between deterministic and random signals versus the total power budget over $L$ frames in sensing-only scenarios.}
        \label{SenOnly_IncreasingL_TotalPower}
\end{figure} 
%\vspace{-0.5em}

In Fig. \ref{SenOnly_IncreasingL_TotalPower}, we present the estimation error results for increasing frame length to illustrate the deterministic-random tradeoff. The total power budget is increased over $L$ frames, with $N_T = 32$. Note the sensing performance mainly depends on the randomness of ISAC signals and the total transmit power. Therefore, for a fixed power budget, the factor impacting the sensing performance is solely the frame length $L$, which correspons to the randomness of ISAC signals. All the curves related to Gaussian signals are generated using the MB-SGP algorithm. It is observed that as the frame length increases, the average sensing performance approaches the LMMSE performance achieved with deterministic signals. This is because the randomness of the signals decreases with longer frame lengths. However, there always is a performance gap between using deterministic and random signals, as indicated by Jensen's inequality. The lessons learned for practical ISAC system design is that one should take the signal randomness into account, especially when the ISAC-BS is unable to accumulate sufficiently long frame length due to time-delay requirement, as well as due to the limited computation, storage, and signal processing capabilities.

\subsection{ISAC Precoding}
In this subsection, our aim is to show the performance tradeoffs between S\&C under different precoding schemes proposed in Sec. \ref{Sec_ISAC}. The benchmark technique, referred to as ``DetOpt'' does not account for the impact of random signals on the sensing performance. Instead, it is based on solving the following deterministic optimization problem 
\begin{figure}[!t]
	\centering
	\includegraphics[width = 0.43\textwidth]{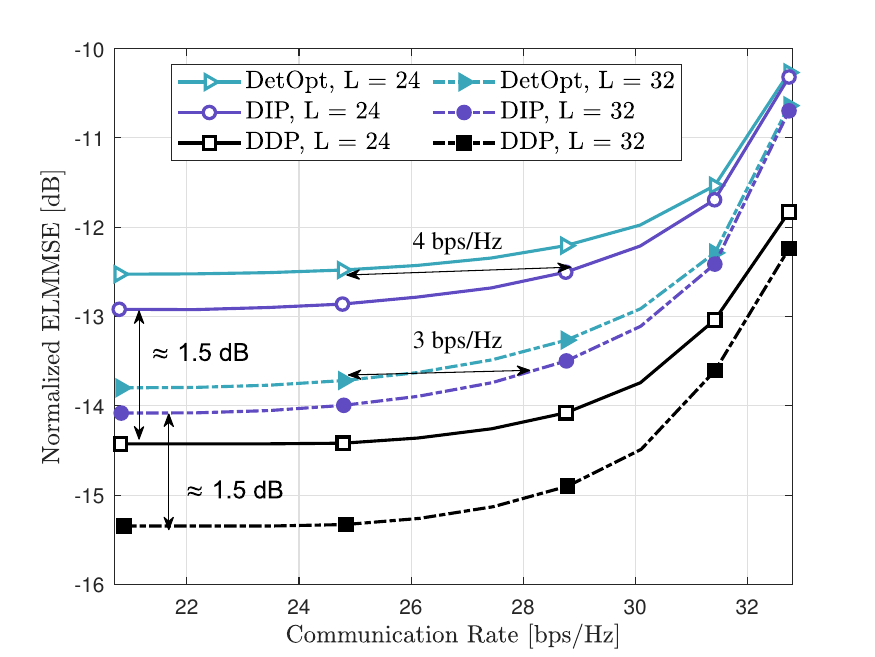}
	\caption{S\&C tradeoff under different precoding schemes with $N_T = N_R = 32$, SNR = $16$ dB.}
        \label{Tradeoff_16dB}
\end{figure} 
\begin{align}
  \min_{\bm{W}} ~ {J}_{\mathsf{LMMSE}} ~~\mathrm{s.t.} ~ R(\bm{W}) \ge R_0, \tr(\bm{WW}^H) \le P.
\end{align}
By denoting the solution by $\bm{W}_{\mathsf{D}}$, the attained ELMMSE of ``DetOpt'' in our simulations is calculated by
\begin{align}
    {J}_{\mathsf{ELMMSE}}^{\mathsf{DetOpt}} = \lim_{N\to\infty}\frac{1}{N} \sum_{n=0}^{N-1} f(\bm{W}_{\mathsf{D}},\bm{S}_n).
\end{align}

In Fig. \ref{Tradeoff_16dB}, we show the performance tradeoff between S\&C by comparing the above benchmark ``DetOpt'' with a pair of precoding schemes, namely, ``DDP'' and ``DIP'' proposed in Sec. \ref{Sec_ISAC} for ISAC precoding. It is observed that our proposed DDP and DIP schemes outperform the benchmark scheme which disregards the randomness of ISAC signals. Specifically, our proposed ``DDP'' scheme exhibits superior performance, i.e., it achieves nearly $1.5$ dB improvement of the sensing performance, while satisfying the required communication rate. Within the same sensing performance, our proposed ``DIP'' scheme acquires $3$ bps/Hz and $4$ bps/Hz communication rate improvement in both cases of $L = 32 = N_T$ and $L = 24 < N_T$, as compared with the benchmark. Fig. \ref{Tradeoff_16dB} also illustrates that the sensing performance deteriorates with the increased rate requirements, as the precoding matrix has to be optimized to meet more stringent communication performance requirements. When the rate requirement is set as its maximum, the precoding matrix approaches the communication-optimal water-filling solution. Consequently, our proposed ``DIP'' scheme and the ``DetOpt'' benchmark exhibit the same worst-case sensing performance as illustrated in Fig. \ref{Tradeoff_16dB}.

\begin{figure}[!t]
	\centering
	\includegraphics[width = 0.41\textwidth]{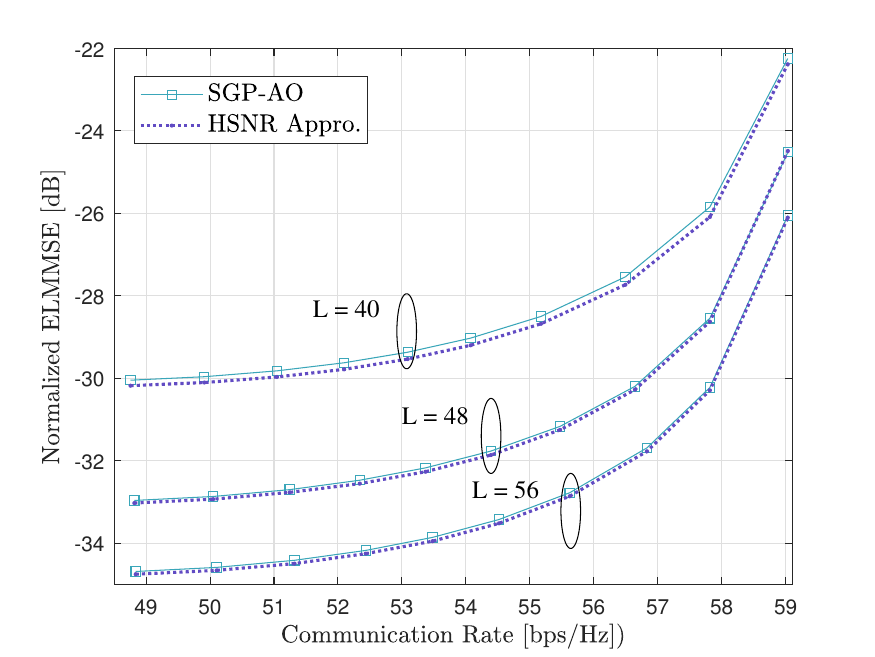}
	\caption{High-SNR regime approximation in an ISAC scenario with $N_T = 32$, SNR = $36$ dB.}
        \label{HSNR_Appro_ISAC}
\end{figure}

In Fig. \ref{HSNR_Appro_ISAC}, we indicate the effectiveness of our proposed SCA algorithm by leveraging the high-SNR approximation (termed as ``HSNR Appro.''). It is observed that as the cumulative frame count increases, our proposed high-SNR approximation-based SCA algorithm closely aligns with the established penalty-based SGP-AO algorithm. This remarkable result suggests that in the high-SNR regime, there is no need to solve stochastic optimization problem \eqref{ELMMSE_Rate_SGP_Op}. Instead, one may utilize the well-established SCA algorithm commonly to solve the asymptotic deterministic optimization problem by leveraging the approximation \eqref{ELMMSE_HSNR_Appro} in Lemma 2. As mentioned earlier, high SNR can compensate for the sensing performance loss caused by signal randomness. In such cases, one may also consider employing precoding schemes for deterministic signals directly due to negligible performance loss.

\begin{figure}[!t]
	\centering
	\includegraphics[width = 0.41\textwidth]{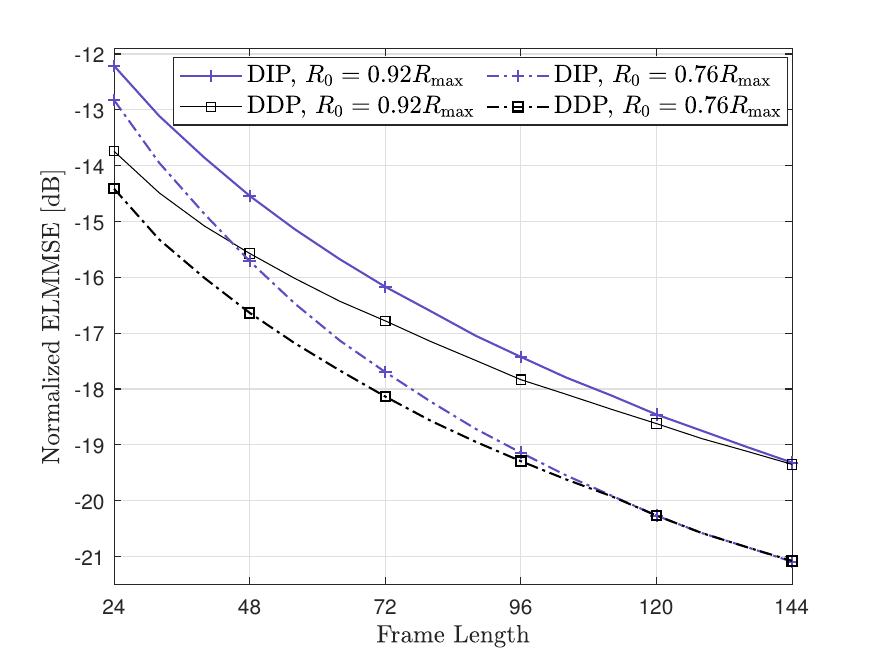}
	\caption{Performance comparison between our proposed DDP and DIP schemes versus the frame length $L$ in ISAC scenario.}
        \label{ISAC_IncreasingL_16dB}
\end{figure}

Finally, we compare the performance of our proposed DDP and DIP schemes tailored for random ISAC signals with an increasing frame length in Fig. \ref{ISAC_IncreasingL_16dB}, where the transmit SNR is set as $16$ dB. It is observed that higher rate threshold constraints result in sensing performance degradation. With a given communication requirement, the sensing performance is improved with increasing accumulated frames. Furthermore, both DDP and DIP schemes asymptotically attain consistent performance as the frame length $L$ increases. This is because the randomness of the ISAC signals becomes progressively negligible as $L$ increases. To this end, the DDP scheme that considers the specific realization of data may no longer significantly boost the sensing performance, which indicates that one may utilize a data-independent precoder with low complexity in practical ISAC systems.

\section{Conclusion}\label{Sec_VI}
In this paper, we investigated dedicated precoding schemes for random ISAC signals in a multi-antenna system. We commenced with an investigation on sensing using random ISAC signals and defined a new sensing metric ELMMSE to characterize the estimation error averaged over the signal randomness. We then proposed a DDP scheme for sensing-only scenarios, which yields a closed-form solution that minimizes the ELMMSE. To reduce the computational complexity of DDP, we further proposed a DIP scheme by the SGP algorithm with a favorable tradeoff between complexity and performance. To gain more design insights, we revealed the optimal structure of the precoding matrix in the high-SNR regime and proposed an SCA algorithm to attain a near-optimal solution. As a step further, we extended both DDP and DIP frameworks into ISAC scenarios by explicitly imposing constraints on the achievable communication rate in the context of ELMMSE minimization problems, which were subsequently solved using a tailored penalty-based alternating optimization algorithm. Finally, we provided extensive simulation results to illustrate the superiority of our proposed dedicated precoding schemes for random ISAC signals.

The proposed precoding methods may be applied to various ISAC systems including mobile perceptive networks and massive-MIMO ISAC systems. In these scenarios, sensing receivers typically cannot accumulate sufficient frame lengths, and the randomness of the ISAC signals cannot be ignored. The future work directions may consider profiling other radar sensing performance metrics under random ISAC signaling, e.g., ergodic CRB, ambiguity function, and mutual information. Moreover, it is necessary to extend this work into such a scenario where the ISAC-BS has no prior knowledge of the sensing channel matrix.

\appendix

\subsection{Proof of Lemma 2}\label{Lemma_2}
Recall that $\bm{R}_H = \bm{Q}\bm{\varLambda}\bm{Q}^H$ denotes the EVD of $\bm{R}_H$. Then, we have
\begin{align}\label{Reform_LMMSE}
 &\mathbb{E}_{\bm{S}} \Big\{\mathrm{tr}\big[\big(\bm{\varLambda}^{-1} + \frac{1}{\sigma_s^2 N_R} \bm{Q}^H \bm{W}\bm{S}\bm{S}^H\bm{W}^H\bm{Q}\big)^{-1}\big] \Big\} \nonumber \\
    & \overset{(a)}{=} \mathbb{E}_{\bm{S}} \Big\{\mathrm{tr}\big[\bm{\varLambda}(\bm{I} + \frac{1}{\sigma_s^2 N_R} \bm{\varLambda}^{\frac{1}{2}}\bm{Q}^H \bm{W}\bm{S}\bm{S}^H\bm{W}^H\bm{Q}\bm{\varLambda}^{\frac{1}{2}})^{-1}\big] \Big\} \nonumber \\
    & \overset{(b)}{=} \gamma\mathbb{E}_{\bm{S}} \left\{\mathrm{tr}\left[\bm{\varLambda}(  \gamma \bm{I} + \bm{R}_{S})^{-1}\right] \right\} \nonumber \\
    & \overset{(c)}{=} \gamma\mathrm{tr}\left\{ \bm{\varLambda} \mathbb{E}_{\bm{S}} \left[(  \gamma \bm{I} + \bm{R}_{S})^{-1}\right] \right\} ,
\end{align}
where $\bm{R}_{S} = (1/P)\bm{\varLambda}^{\frac{1}{2}}\bm{Q}^H \bm{W}\bm{S}\bm{S}^H\bm{W}^H\bm{Q}\bm{\varLambda}^{\frac{1}{2}}$ and $\gamma= \sigma_s^2 N_R/{P}$. Here, $(a)$ and $(b)$ hold due to the property of the matrix trace, and $(c)$ holds due to the linear property of the matrix trace and expectation. Furthermore, the expectation of $( \gamma \bm{I} + \bm{R}_{S})^{-1}$ can be approximated by the second-order expansion of the von Neumann series, expressed as
\begin{align}\label{Matrix_Expectation_appro}
    \mathbb{E}_{\bm{S}} \left[(  \gamma \bm{I} + \bm{R}_{S})^{-1}\right] \approx  \mathbb{E}\left[\bm{R}_{S}^{-1}\right]-\gamma\mathbb{E}\left[\bm{R}_{S}^{-2}\right]+\mathcal{O}(\gamma^2),
\end{align}
which is only valid for very small $\gamma$. In a practical scenario, one can easily implement a high transmit power $P$ to get small $\gamma$ with given $N_R$ and $\sigma_s^2$, which may achieve an accurate approximation in a high-SNR regime. Therefore, the asymptotic formulation of \eqref{Reform_LMMSE} is calculated by 
\begin{align}\label{fW_Matrix}
    \mathbb{E}_{\bm{S}}[f(\bm{W} ; \bm{S})] \approx \gamma\mathrm{tr}\left\{\bm{\varLambda}\mathbb{E}\left[\bm{R}_{S}^{-1}\right]-\gamma\bm{\varLambda} \mathbb{E}\left[\bm{R}_{S}^{-2}\right] \right\}. 
\end{align} 

We observe that $\bm{R}_{S}^{-1}$ follows the complex inverse Wishart distribution with $L$ degrees of freedom and parameter matrix $\bm{\Sigma} = (1/P)\bm{\varLambda}^{\frac{1}{2}}\bm{Q}^H \bm{W}\bm{W}^H\bm{Q}\bm{\varLambda}^{\frac{1}{2}}$. 
The first and second moments of the complex inverse Wishart matrix are given by \cite{Wishart_Moment}
\begin{subequations}\label{IWishart_Moments}
    \begin{align}
    \mathbb{E}\left[\bm{R}_{S}^{-1}\right] &= \frac{1}{L-N_T}\bm{\Sigma}^{-1}, \\
    % \mathbb{E}\left[\bm{R}_{S}^{-2}\right] & = \frac{1}{(L-N_T)^2}\bm{\Sigma}^{-2} + \frac{1}{(L-N_T)^3}\mathrm{tr}(\bm{\Sigma}^{-1})\bm{\Sigma}^{-1}, 
    \mathbb{E}\left[\bm{R}_{S}^{-2}\right] &= \frac{1}{(L-N_T)^2}\bm{\Sigma}^{-2} + \frac{1}{(L-N_T)^3}\mathrm{tr}(\bm{\Sigma}^{-1})\bm{\Sigma}^{-1} \nonumber\\
    &\quad+\underbrace{\mathcal{O}\left(\frac{1}{(L-N_T)^4}\right)\bm{I}_{N_T}}_{\mathrm{error~term}} .
\end{align}
\end{subequations}
Note that \eqref{IWishart_Moments} is valid when $(L-N_T)^4$ is sufficiently large, primarily due to the error term. Consequently, by substituting \eqref{IWishart_Moments} into \eqref{fW_Matrix} and performing some algebraic operations, we obtain an asymptotic expression in the high-SNR regime as shown in \eqref{ELMMSE_HSNR_Appro}, thereby concluding the proof.

\subsection{Proof of Theorem 2}\label{Theorem_2}
Let $\bm{R} = \bm{Q}\bm{\varOmega }\bm{Q}^H$. Since $\bm{Q}$ is a unitary matrix, we have 
\begin{subequations}
\begin{align}
    \tilde{f}(\bm{\varOmega }) &= \tilde{f}_1(\bm{R}) + \tilde{f}_2(\bm{R}), \\
    \tr(\bm{\varOmega }) &= \tr(\bm{R}),
\end{align}
\end{subequations}
where $\tilde{f}_1(\bm{R}) = \kappa_1 \mathrm{tr}(\bm{R}^{-1})$ and $\tilde{f}_2(\bm{R}) =  - \kappa_2 \mathrm{tr}(\bm{\varLambda}^{-1} \bm{R}^{-2} )- \kappa_3 \mathrm{tr}(\bm{R}^{-1} ) \mathrm{tr}(\bm{\varLambda}^{-1}\bm{R}^{-1})$.
Therefore, problem \eqref{LMMSE_ApproxProblem} is recast as 
\begin{align}\label{LMMSE_RWhat}
        \min_{\bm{R}} ~ \tilde{f}_1(\bm{R}) + \tilde{f}_2(\bm{R})
		~~ \mathrm{s.t.} ~ \tr(\bm{R}) \le P, \bm{R} \succeq \bm{0}.
\end{align}
It is obvious that problem \eqref{LMMSE_RWhat} is equivalent to \eqref{LMMSE_ApproxProblem}. 

Let the elements of $\bm{\varLambda}$ be sorted in the increasing order such that the elements of $\bm{\varLambda}^{-1}$ are sorted in the decreasing order, denoted by $\bm{\varLambda}^{-1} = \mathrm{diag}([\hat{\lambda}_1,\hat{\lambda}_2,\dots,\hat{\lambda}_{N_T} ]) $,  $\hat{\lambda}_1 \ge \hat{\lambda}_2 \ge\dots \ge \hat{\lambda}_{N_T}$. By applying Lemma 3, we have 
\begin{subequations}
\begin{align}
        \mathrm{tr}(\bm{\varLambda}^{-1} \bm{R}^{-2} ) \le \mathrm{tr}(\bm{\varLambda}^{-1} \bm{R}^{-2}_{\mathsf{opt}} ) , \\
         \mathrm{tr}(\bm{\varLambda}^{-1} \bm{R}^{-1} ) \le \mathrm{tr}(\bm{\varLambda}^{-1} \bm{R}^{-1}_{\mathsf{opt}} ).
\end{align}
\end{subequations}
Here, the equality holds if and only if $\bm{R}_{\mathsf{opt}}^{-1}$ is a diagonal matrix with elements sorted in the decreasing order, yielding
\begin{subequations}
\begin{align}
        \bm{R}_{\mathsf{opt}}^{-1} =  \mathrm{diag}([p_1,\dots,p_{N_T} ]),~p_1 \ge \dots \ge p_{N_T}, \\
         \bm{R}_{\mathsf{opt}}^{-2} =  \mathrm{diag}([p_1^2,\dots,p_{N_T}^2 ]),~p_1^2 \ge \dots \ge p_{N_T}^2.
\end{align}
\end{subequations}
Due to $\kappa_2 > 0$ and $\kappa_3 > 0$, the objective function of \eqref{LMMSE_RWhat} attains its lower bound, expressed by
\begin{align}
        \tilde{f}_1(\bm{R}) + \tilde{f}_2(\bm{R})  \ge   \tilde{f}_1(\bm{R}_{\mathsf{opt}}) + \tilde{f}_2(\bm{R}_{\mathsf{opt}}). 
        % \kappa_1 \mathrm{tr}(\bm{R}_{\mathsf{opt}}^{-1}) - \kappa_2 \mathrm{tr}(\bm{\varLambda}^{-1} \bm{R}_{\mathsf{opt}}^{-2} )
        % - \kappa_3 \mathrm{tr}(\bm{R}_{\mathsf{opt}}^{-1})  \\ \times \mathrm{tr}(\bm{\varLambda}^{-1}\bm{R}_{\mathsf{opt}}^{-1}). 
\end{align}
% \begin{multline}
%         \tilde{f}_1(\bm{R}) + \tilde{f}_2(\bm{R})  \ge  \kappa_1 \mathrm{tr}(\bm{R}_{\mathsf{opt}}^{-1}) - \kappa_2 \mathrm{tr}(\bm{\varLambda}^{-1} \bm{R}_{\mathsf{opt}}^{-2} )
%         - \kappa_3 \mathrm{tr}(\bm{R}_{\mathsf{opt}}^{-1})  \\ \times \mathrm{tr}(\bm{\varLambda}^{-1}\bm{R}_{\mathsf{opt}}^{-1}). 
% \end{multline}
Therefore, we can obtain the optimal solution $\bm{\varOmega }^{\mathsf{opt}}$ of problem \eqref{LMMSE_RWhat}, expressed as
\begin{align}
    \bm{\varOmega }^{\mathsf{opt}} = \bm{Q}^H \bm{R}_{\mathsf{opt}} \bm{Q} \triangleq \bm{Q}^H \bm{\varPsi}^{\uparrow} \bm{Q},
\end{align}
where $\bm{\varPsi}^{\uparrow} = \mathrm{diag}([\psi_1,\psi_2,\dots,\psi_{N_T} ])$, $\psi_1 \le \psi_2 \le \dots \le \psi_{N_T}$, and $\psi_{\ell} = 1/p_{N_T+1-\ell},~\ell = 1,2, \dots, N_T$, completing the proof.

\subsection{Proof of Proposition 2}\label{Proposition_2}
Let us consider the Gaussian linear channel model
\begin{align}
    \bm{Y}_c = \bm{H}_c \bm{W} \bm{S} + \bm{Z}_s.
\end{align}
The mutual information between $\bm{Y}_c$ and $\bm{X} = \bm{WS}$ conditioned on $\bm{H}_c$ is lower-bounded by
\begin{align}\label{MI}
    & I(\bm{Y}_c; \bm{WS} | \bm{H}_c) =  I(\bm{Y}_c; \bm{W}, \bm{S} | \bm{H}_c) \nonumber \\
    &\overset{(a)}{=} I(\bm{Y}_c; \bm{W} | \bm{H}_c) + I(\bm{Y}_c; \bm{S} | \bm{H}_c, \bm{W}) \nonumber \\
    & \overset{(b)}{\ge} I(\bm{Y}_c; \bm{S} | \bm{H}_c, \bm{W}).
\end{align}
where (a) holds due to the chain rule for mutual information and (b) holds due to the fact $I(\bm{Y}_c; \bm{W} | \bm{H}_c) \ge 0$. Denote the channel capacity as $C$ and take the maximum on both sides of \eqref{MI} over all possible input distributions, yielding
\begin{align}
    C &= \max I(\bm{Y}_c; \bm{WS} | \bm{H}_c) \ge \max  I(\bm{Y}_c; \bm{S} | \bm{H}_c, \bm{W}) \nonumber \\
    &\overset{(c)}{\ge} \max  I(\bm{Y}_c; \bm{S} | \bm{H}_c, \bm{W})~ \mathrm{s.t.}~ \bm{W} \perp \!\!\! \perp \bm{S} \nonumber \\
    &\overset{(d)}{=} \mathbb{E}_{\bm{W}} \big[ \log \det(  \bm{I}_{N_u} + \sigma_c^{-2} \bm{H}_{c} \bm{W}  \bm{W}^{H}\bm{H}_{c}^{H} )\big],
\end{align}
where (c) holds due to the extra constraint that $\bm{W}$ and $\bm{S}$ are independent, denoted by $\bm{W} \perp \!\!\! \perp \bm{S}$. Moreover, (d) holds since $\bm{H}_c$ is perfectly known at the transmitter, completing the proof.

\bibliographystyle{IEEEtran}
\bibliography{ref}

\begin{thebibliography}{10}
\providecommand{\url}[1]{#1}
\csname url@samestyle\endcsname
\providecommand{\newblock}{\relax}
\providecommand{\bibinfo}[2]{#2}
\providecommand{\BIBentrySTDinterwordspacing}{\spaceskip=0pt\relax}
\providecommand{\BIBentryALTinterwordstretchfactor}{4}
\providecommand{\BIBentryALTinterwordspacing}{\spaceskip=\fontdimen2\font plus
\BIBentryALTinterwordstretchfactor\fontdimen3\font minus
  \fontdimen4\font\relax}
\providecommand{\BIBforeignlanguage}[2]{{%
\expandafter\ifx\csname l@#1\endcsname\relax
\typeout{** WARNING: IEEEtran.bst: No hyphenation pattern has been}%
\typeout{** loaded for the language `#1'. Using the pattern for}%
\typeout{** the default language instead.}%
\else
\language=\csname l@#1\endcsname
\fi
#2}}
\providecommand{\BIBdecl}{\relax}
\BIBdecl

\bibitem{lushihang2023sensing}
S.~Lu, F.~Liu, F.~Dong, Y.~Xiong, J.~Xu, and Y.-F. Liu, ``Sensing with random
  signals,'' \emph{arXiv preprint arXiv:2309.02375}, 2023.

\bibitem{cui2021integrating}
Y.~Cui, F.~Liu, X.~Jing, and J.~Mu, ``Integrating sensing and communications
  for ubiquitous {IoT}: {Applications}, trends, and challenges,'' \emph{IEEE
  Network}, vol.~35, no.~5, pp. 158--167, Sept. 2021.

\bibitem{Chafii2023CST}
M.~Chafii, L.~Bariah, S.~Muhaidat, and M.~Debbah, ``Twelve scientific
  challenges for {6G: Rethinking} the foundations of communications theory,''
  \emph{IEEE Commun. Surveys Tuts.}, vol.~25, no.~2, pp. 868--904,
  Secondquarter 2023.

\bibitem{liuan2022survey}
A.~Liu, Z.~Huang, M.~Li, Y.~Wan, W.~Li, T.~X. Han, C.~Liu, R.~Du, D.~K.~P. Tan,
  J.~Lu, Y.~Shen, F.~Colone, and K.~Chetty, ``A survey on fundamental limits of
  integrated sensing and communication,'' \emph{IEEE Commun. Surveys Tuts.},
  vol.~24, no.~2, pp. 994--1034, 2nd quarter 2022.

\bibitem{liufan2023seventy}
F.~Liu, L.~Zheng, Y.~Cui, C.~Masouros, A.~P. Petropulu, H.~Griffiths, and Y.~C.
  Eldar, ``Seventy years of radar and communications: {T}he road from
  separation to integration,'' \emph{IEEE Signal Process. Mag.}, vol.~40,
  no.~5, pp. 106--121, Jul. 2023.

\bibitem{ITU2023}
{ITU-R WP5D}, ``{D}raft {N}ew {Recommendation ITU-R M. [IMT. FRAMEWORK FOR 2030
  AND BEYOND]},'' 2023.

\bibitem{liuxiangTSP2020}
X.~Liu, T.~Huang, N.~Shlezinger, Y.~Liu, J.~Zhou, and Y.~C. Eldar, ``{Joint
  transmit beamforming for multiuser MIMO communications and MIMO radar},''
  \emph{IEEE Trans. Signal Process.}, vol.~68, pp. 3929--3944, Jun. 2020.

\bibitem{liufanTSP2018mimo}
F.~Liu, C.~Masouros, A.~Li, T.~Ratnarajah, and J.~Zhou, ``{MIMO radar and
  cellular coexistence: A power-efficient approach enabled by interference
  exploitation},'' \emph{IEEE Trans. Signal Process.}, vol.~66, no.~14, pp.
  3681--3695, Jul. 2018.

\bibitem{chenliTSP2021joint}
L.~Chen, F.~Liu, W.~Wang, and C.~Masouros, ``Joint radar-communication
  transmission: {A generalized Pareto optimization framework},'' \emph{IEEE
  Trans. Signal Process.}, vol.~69, pp. 2752--2765, May 2021.

\bibitem{Nguyen2023TSP}
N.~T. Nguyen, N.~Shlezinger, Y.~C. Eldar, and M.~Juntti, ``Multiuser {MIMO}
  wideband joint communications and sensing system with subcarrier
  allocation,'' \emph{IEEE Trans. Signal Process.}, vol.~71, pp. 2997--3013,
  Aug. 2023.

\bibitem{ren2023fundamental}
Z.~Ren, Y.~Peng, X.~Song, Y.~Fang, L.~Qiu, L.~Liu, D.~W.~K. Ng, and J.~Xu,
  ``Fundamental {CRB}-rate tradeoff in multi-antenna {ISAC} systems with
  information multicasting and multi-target sensing,'' \emph{IEEE Trans.
  Wireless Commun.}, pp. 1--1, 2023.

\bibitem{MarwaTWC2023}
A.~Bazzi and M.~Chafii, ``On outage-based beamforming design for
  dual-functional radar-communication {6G} systems,'' \emph{IEEE Trans.
  Wireless Commun.}, vol.~22, no.~8, pp. 5598--5612, Aug 2023.

\bibitem{Huahaocheng2023TWC}
H.~Hua, T.~X. Han, and J.~Xu, ``{MIMO} integrated sensing and communication:
  {CRB}-rate tradeoff,'' \emph{IEEE Trans. Wireless Commun.}, Aug. 2023,
  doi:10.1109/TWC.2023.3303326.

\bibitem{kobayashiTIT2022}
M.~Ahmadipour, M.~Kobayashi, M.~Wigger, and G.~Caire, ``An
  information-theoretic approach to joint sensing and communication,''
  \emph{IEEE Trans. Inf. Theory}, pp. 1--1, 2022.

\bibitem{xiong2023fundamental}
Y.~Xiong, F.~Liu, Y.~Cui, W.~Yuan, T.~X. Han, and G.~Caire, ``On the
  fundamental tradeoff of integrated sensing and communications under
  {G}aussian channels,'' \emph{IEEE Trans. Inf. Theory}, vol.~69, no.~9, pp.
  5723--5751, Jun. 2023.

\bibitem{mishra2019SPM}
K.~V. Mishra, M.~B. Shankar, V.~Koivunen, B.~Ottersten, and S.~A. Vorobyov,
  ``Toward millimeter-wave joint radar communications: {A} signal processing
  perspective,'' \emph{IEEE Signal Process. Mag.}, vol.~36, no.~5, pp.
  100--114, Sept. 2019.

\bibitem{keskin2021TSPlimitedfeedback}
M.~F. Keskin, V.~Koivunen, and H.~Wymeersch, ``Limited feedforward waveform
  design for {OFDM} dual-functional radar-communications,'' \emph{IEEE Trans.
  Signal Process.}, vol.~69, pp. 2955--2970, Apr. 2021.

\bibitem{chengziyang2021hybrid}
Z.~Cheng, Z.~He, and B.~Liao, ``Hybrid beamforming design for {OFDM}
  dual-function radar-communication system,'' \emph{IEEE J. Sel. Topics Signal
  Process.}, vol.~15, no.~6, pp. 1455--1467, Oct. 2021.

\bibitem{muxidong2022JSAC}
X.~Mu, Y.~Liu, L.~Guo, J.~Lin, and L.~Hanzo, ``{NOMA}-aided joint radar and
  multicast-unicast communication systems,'' \emph{IEEE J. Sel. Areas Commun.},
  vol.~40, no.~6, pp. 1978--1992, Mar. 2022.

\bibitem{Sankar2023twc}
R.~P. Sankar, S.~P. Chepuri, and Y.~C. Eldar, ``Beamforming in integrated
  sensing and communication systems with reconfigurable intelligent surfaces,''
  \emph{IEEE Trans. Wireless Commun.}, 2023, doi:10.1109/TWC.2023.3313938.

\bibitem{liufan2018TSP}
F.~Liu, L.~Zhou, C.~Masouros, A.~Li, W.~Luo, and A.~Petropulu, ``Toward
  dual-functional radar-communication systems: {O}ptimal waveform design,''
  \emph{IEEE Trans. Signal Process.}, vol.~66, no.~16, pp. 4264--4279, Jun.
  2018.

\bibitem{liu2021cramer}
F.~Liu, Y.-F. Liu, A.~Li, C.~Masouros, and Y.~C. Eldar, ``Cram{\'e}r-{Rao}
  bound optimization for joint radar-communication beamforming,'' \emph{IEEE
  Trans. Signal Process.}, vol.~70, pp. 240--253, Dec. 2021.

\bibitem{songxianxin2023TSP}
X.~Song, J.~Xu, F.~Liu, T.~X. Han, and Y.~C. Eldar, ``Intelligent reflecting
  surface enabled sensing: {Cramér-Rao} bound optimization,'' \emph{IEEE
  Trans. Signal Process.}, vol.~71, pp. 2011--2026, May 2023.

\bibitem{mengkt2023TWC}
K.~Meng, Q.~Wu, S.~Ma, W.~Chen, K.~Wang, and J.~Li, ``Throughput maximization
  for {UAV}-enabled integrated periodic sensing and communication,'' \emph{IEEE
  Trans. Wireless Commun.}, vol.~22, no.~1, pp. 671--687, Aug. 2023.

\bibitem{liurang2021JSTSP}
R.~Liu, M.~Li, Q.~Liu, and A.~L. Swindlehurst, ``Dual-functional
  radar-communication waveform design: {A} symbol-level precoding approach,''
  \emph{IEEE J. Sel. Topics Signal Process.}, vol.~15, no.~6, pp. 1316--1331,
  Sept. 2021.

\bibitem{wei20225g}
Z.~Wei, Y.~Wang, L.~Ma, S.~Yang, Z.~Feng, C.~Pan, Q.~Zhang, Y.~Wang, H.~Wu, and
  P.~Zhang, ``5g prs-based sensing: A sensing reference signal approach for
  joint sensing and communication system,'' \emph{IEEE Trans. Veh. Technol.},
  vol.~72, no.~3, pp. 3250--3263, Oct. 2022.

\bibitem{lin20215g}
X.~Lin and N.~Lee, ``5g and beyond,'' \emph{Cham, Switzerland: Springer Nature
  Switzerland AG}, 2021.

\bibitem{zhangyumeng2023input}
Y.~Zhang, S.~Aditya, and B.~Clerckx, ``Input distribution optimization in
  {OFDM} dual-function radar-communication systems,'' \emph{arXiv preprint
  arXiv:2305.06635}, 2023.

\bibitem{xiong2023generalized}
Y.~Xiong, F.~Liu, and M.~Lops, ``Generalized deterministic-random tradeoff in
  integrated sensing and communications: {The} sensing-optimal operating
  point,'' \emph{arXiv preprint arXiv:2308.14336}, 2023.

\bibitem{cover1999elements}
T.~M. Cover, \emph{Elements of {Information Theory}}.\hskip 1em plus 0.5em
  minus 0.4em\relax John Wiley \& Sons, 1999.

\bibitem{stoica2007probing}
P.~Stoica, J.~Li, and Y.~Xie, ``On probing signal design for {MIMO} radar,''
  \emph{IEEE Trans. Signal Process.}, vol.~55, no.~8, pp. 4151--4161, May 2007.

\bibitem{bekkerman2006target}
I.~Bekkerman and J.~Tabrikian, ``Target detection and localization using {MIMO}
  radars and sonars,'' \emph{IEEE Trans. Signal Process.}, vol.~54, no.~10, pp.
  3873--3883, Sept. 2006.

\bibitem{du2023reshaping}
Z.~Du, F.~Liu, Y.~Xiong, T.~X. Han, Y.~C. Eldar, and S.~Jin, ``Reshaping the
  {ISAC} tradeoff under {OFDM} signaling: {A} probabilistic constellation
  shaping approach,'' \emph{arXiv preprint arXiv:2312.15941}, 2023.

\bibitem{lu2022performance}
S.~Lu, X.~Meng, Z.~Du, Y.~Xiong, and F.~Liu, ``On the performance gain of
  integrated sensing and communications: {A} subspace correlation
  perspective,'' in \emph{Proc. IEEE ICC}, Jul. 2023, pp. 1--6.

\bibitem{liu_Yafeng2022joint}
F.~Liu, Y.-F. Liu, C.~Masouros, A.~Li, and Y.~C. Eldar, ``A joint
  radar-communication precoding design based on {C}ram{\'e}r-{Rao} bound
  optimization,'' in \emph{Proc. IEEE Radar Conf. (RadarConf22)}, May 2022, pp.
  1--6.

\bibitem{tangbo2019TSP}
B.~Tang and J.~Li, ``Spectrally constrained {MIMO} radar waveform design based
  on mutual information,'' \emph{IEEE Trans. Signal Process.}, vol.~67, no.~3,
  pp. 821--834, Dec. 2019.

\bibitem{tangbo2010TSP}
B.~Tang, J.~Tang, and Y.~Peng, ``{MIMO} radar waveform design in colored noise
  based on information theory,'' \emph{IEEE Trans. Signal Process.}, vol.~58,
  no.~9, pp. 4684--4697, May 2010.

\bibitem{boyd2004convex}
S.~P. Boyd and L.~Vandenberghe, \emph{Convex {Optimization}}.\hskip 1em plus
  0.5em minus 0.4em\relax {Cambridge University Press}, 2004.

\bibitem{biguesh2006training}
M.~Biguesh and A.~B. Gershman, ``Training-based {MIMO} channel estimation: {A}
  study of estimator tradeoffs and optimal training signals,'' \emph{IEEE
  Trans. Signal Process.}, vol.~54, no.~3, pp. 884--893, Feb. 2006.

\bibitem{tang2011waveform}
B.~Tang, J.~Tang, and Y.~Peng, ``Waveform optimization for {MIMO} radar in
  colored noise: {Further} results for estimation-oriented criteria,''
  \emph{IEEE Trans. Signal Process.}, vol.~60, no.~3, pp. 1517--1522, Nov.
  2011.

\bibitem{liuan2019stochastic}
A.~Liu, V.~K. Lau, and B.~Kananian, ``Stochastic successive convex
  approximation for non-convex constrained stochastic optimization,''
  \emph{IEEE Trans. Signal Process.}, vol.~67, no.~16, pp. 4189--4203, Jul.
  2019.

\bibitem{li2019convergence}
X.~Li and F.~Orabona, ``On the convergence of stochastic gradient descent with
  adaptive stepsizes,'' in \emph{Proc. 22nd Int. Conf. Artif. Intell.
  Stat.}\hskip 1em plus 0.5em minus 0.4em\relax PMLR, Apr. 2019, pp. 983--992.

\bibitem{kingma2014adam}
D.~P. Kingma and J.~Ba, ``Adam: {A} method for stochastic optimization,''
  \emph{Proc. Int. Conf. Learn. Represent.}, pp. 1--15, 2015.

\bibitem{book2021kkt}
B.~Ghojogh, A.~Ghodsi, F.~Karray, and M.~Crowley, ``Kkt conditions, first-order
  and second-order optimization, and distributed optimization: Tutorial and
  survey,'' \emph{arXiv preprint arXiv:2110.01858}, 2021.

\bibitem{marshall1979inequalities}
A.~W. Marshall, I.~Olkin, and B.~C. Arnold, \emph{Inequalities: {Theory of
  Majorization and its Applications}}.\hskip 1em plus 0.5em minus 0.4em\relax
  Springer, 1979.

\bibitem{Wishart_Moment}
D.~Maiwald and D.~Kraus, ``On moments of complex {Wishart} and complex inverse
  {Wishart} distributed matrices,'' in \emph{Proc. IEEE ICASSP}, vol.~5, 1997,
  pp. 3817--3820.

\end{thebibliography}

\end{document}